\documentclass[twocolumn,secnumarabic,amssymb, nobibnotes,superscriptaddress, aps, prd]{revtex4-2}
\usepackage{amsmath,amssymb}
\usepackage{bigints}
\usepackage{graphicx}
\usepackage{multirow}
\usepackage{caption}
\usepackage{subcaption}
\usepackage{orcidlink}

\begin{document}

\title{Gravitational wave echoes as probes of the maximum mass of strange stars in quadratic curvature-matter coupled gravity}%

\author{Debadri Bhattacharjee\orcidlink{0000-0002-7007-9397}}
\email{debadriwork@gmail.com}
\affiliation{IUCAA Centre for Astronomy Research and Development (ICARD), Department of Physics, Cooch Behar Panchanan Barma University, Vivekananda Street, District: Cooch Behar, Pin: 736101, West Bengal, India}

\author{Pradip Kumar Chattopadhyay\orcidlink{0000-0001-6662-8800}}
\email{pkc\_76@rediffmail.com}
\affiliation{IUCAA Centre for Astronomy Research and Development (ICARD), Department of Physics, Cooch Behar Panchanan Barma University, Vivekananda Street, District: Cooch Behar, Pin: 736101, West Bengal, India}

\author{Kazuharu Bamba \orcidlink{0000-0001-9720-8817}}
\email{bamba@sss.fukushima-u.ac.jp}
\affiliation{Division of Human Support System, Faculty of Symbiotic Systems Science, Fukushima University, Fukushima 960-1296, Japan}


\begin{abstract}
	Gravitational wave astronomy provides a powerful avenue for probing exotic compact stars. Recent analyses of GW170817 have reported possible post-merger gravitational wave echoes with a significance of $4.2\sigma$ and a dominant frequency near $72,\text{Hz}$. Such echoes may arise from ultracompact remnants possessing photon spheres that partially trap gravitational perturbations. In general relativity, photon-sphere formation requires the stellar compactness to lie between one-third and four-ninths, a condition difficult to achieve with realistic equations of state. Here, we investigate this possibility in quadratic curvature gravity with non-minimal matter coupling, considering strange stars described by the MIT bag model equation of state. By solving the modified Tolman-Oppenheimer-Volkoff equations, we obtain mass-radius relations and identify configurations capable of supporting photon spheres and gravitational wave echoes. The modified Buchdahl limit permits more compact stellar solutions, while photon-sphere constraints restrict the viable parameter space. We find that increasing the bag constant reduces the maximum mass and echo time, shifting the echo frequency toward the kHz regime. The echo constraints impose more stringent maximum mass-radius limits than hydrostatic equilibrium, suggesting revised maximum mass bounds for strange stars. We further compute the critical impact parameter and eikonal quasi-normal modes, demonstrating how modified gravity parameters systematically alter the photon-sphere radius and associated observational signatures. These results establish post-merger strange stars as potential GW echo sources and highlight echoes, optical shadows, and QNMs as complementary probes of modified gravity for next-generation detectors such as the Einstein Telescope and Cosmic Explorer.
\end{abstract}
\maketitle

\section{Introduction}\label{sec1}
The resounding success of the LIGO-Virgo-KAGRA (LVK) network in detecting gravitational waves (GW) originating from compact binary coalescence has significantly advanced our understanding of black holes and compact stars \cite{Abbott}. These relativistic mergers have emerged as the Rosetta Stone for exploring gravity in its strongest-field regime. To date, the LVK network has successfully recorded more that 200 compact binary coalescence events \cite{Abbott1,Abbott2,Abbott3,Abac}, among which the recently detected GW250114 is significantly important owing to its exceptionally high ringdown signal-to-noise ratio (SNR). This high SNR has facilitated high-precision black hole spectroscopy \cite{Abac1,Abac2}. The early-time ringdown signal is found to be in good agreement with the predictions of general relativity (GR) for Kerr black holes \cite{Isi,Capano,Cotesta}. However, this phase is expected to exhibit limited sensitivity to modifications in the near-horizon region because the strong gravitational redshift suppresses the observational imprint of such effects until substantially later times.

If, instead of a classical BH, the post-merger remnant is an ultra compact object (UCO) possessing a partially or fully reflective surface located just outside the would-be horizon, as proposed in several quantum-gravity-inspired scenarios, GW can become confined between the angular momentum (light-ring) potential barrier and the inner reflective boundary. The successive leakage of GW through the potential barrier then generates a sequence of damped, nearly periodic wave packets that appear after the primary ringdown signal. These are commonly referred to as GW echoes \cite{Cardoso,Cardoso1,Bueno,Zhang,Urbano,Mannarelli,Li,Dong}. Notably, the echo time delay is observationally attainable even when the reflective surface lies only one Planck scale distance away from the would-be horizon. 

The generation of GW echoes require a massive post-merger object, with a mass $M$, featuring a photon sphere at $R=3M$ in  the framework of GR \cite{Weinberg,Misner,Claudel}. At $R=3M$, circular photon orbits are possible owing to the angular potential barrier. This feature is present in case of black holes \cite{Shapiro} as well as UCOs \cite{Iyer,Nemiroff}. To support a photon sphere and, consequently, GW echoes within GR, neutron stars must attain extremely high compactness to become ultracompact objects. Achieving such compact configurations with realistic neutron star EoS is challenging. In contrast, SSs are hypothesised to consist of deconfined strange quark matter, in which the hadronic degrees of freedom associated with individual nucleons are expected to dissolve into their constituent quark degrees of freedom at sufficiently high densities. This transition to a highly dense quark matter phase can, for suitable EoS, yield more compact stellar configurations than conventional neutron stars. Consequently, SSs may attain compactness (M/R>1/3), placing their stellar surface within the photon sphere at R=3M. Such ultracompact SS configurations may therefore serve as distinct theoretical candidates for GW echoes. Notably, in case of black holes, the emergence of echoes require a second reflecting surface since there is GW absorption due to quantum corrections near the horizon \cite{Ferrari}. On the other hand, for UCOs such internal reflecting surface remains absent since there is no absorption of considerable fraction of GW. 

GW170817 event detected by the LIGO-Virgo collaboration \cite{Abbott4} showed the merging of binary neutron stars (NS) and inferred a system of total mass $2.74_{-0.01}^{+0.04}~\mathrm{M_{\odot}}$. The nature of this final object, whether it is a black hole or a massive compact star remains elusive. Using the post-merger object of GW170817, the aspect of GW echoes was studied in Ref. \cite{Abedi} reporting a frequency of $\approx72~\mathrm{Hz}$ with significance of $4.2\sigma$. This signal was considered to be originating from quantum effects near the black hole horizon. An explanation of this echo from the perspective of UCOs was presented by Pani and Ferrari \cite{Pani}. Interestingly, using an simple form of incompressible equation of state (EoS), they showed that for producing an echo of such low frequency, the compactness of the NS must be very close to the Buchdahl's compactness limit \cite{Buchdahl}, with a radius of $R_{b}=\frac{9M}{4}$. Hence, for UCOs in the framework of GR, the compactness must be, (i) greater than $\frac{1}{3}$ to establish the possibility of a photon sphere, and (ii) less than $\frac{4}{9}$ to produce low frequency (tens of Hz) GW echoes.

The EoS of ultra-dense matter in NS interiors remains an open problem, for which the strange quark matter (SQM) hypothesis provides a viable description. Its predictions are consistent with current observational constraints from massive NS, GW, and radius measurements. Initially, Itoh \cite{Itoh} demonstrated that hypothetical quark stars, composed of approximately equal numbers of $u$, $d$, and $s$ quarks, can attain stable hydrostatic equilibrium without undergoing gravitational collapse. It must be noted that the stability of quark matter is determined by its energy per baryon ($\mathcal{E}_{B}$). While the most stable nucleus, $^{56}\mathrm{Fe}$, has $\mathcal{E}_{B}=930.4~\mathrm{MeV}$, Madsen \cite{Madsen} showed that $ud$-quark matter has $\mathcal{E}_{B}=934~\mathrm{MeV}$ at zero external pressure, rendering it more unstable than strange quark matter. The inclusion of $s$ quarks lowers this value to approximately $829~\mathrm{MeV}$, making SQM energetically favourable and more densely bound. This supports the Bodmer-Witten conjecture \cite{Bodmer,Witten}, according to which SQM represents the absolute ground state of quantum chromodynamics (QCD).

The asymptotic freedom of QCD further predicts that, at sufficiently high densities, hadrons deconfine into quarks and gluons, forming a weakly interacting quark-gluon plasma. Under the high baryon density and relatively low temperature expected in NS cores, deconfined quark matter may therefore exist \cite{Alford}. This possibility has led to the proposal of a distinct class of compact objects known as strange stars (SS) \cite{Alcock,Haensel,Kettner}. Owing to the lack of comprehensive understanding of the hadron-quark phase transition, several phenomenological EoSs have been developed to describe SQM. Among these, the MIT bag model EoS \cite{Kettner} remains one of the most successful, providing a simple yet accurate description of compact stars that is consistent with current astrophysical observations.

The post-merger GW signal may exhibit distinctive features. However, to the best of our knowledge, no numerical simulations have yet investigated NS mergers that culminate in the formation of SS. Consequently, the present work is restricted to an analysis of the post-merger GW echo signal. Notably, the investigations of GW echoes originating from SS and exotic compact stars have been reported in Refs. \cite{Urbano,Mannarelli,Zhang1,Bora}. Furthermore, to produce echoes, the compactness must lie within the previously mentioned limits to achieve the notion of photon spheres. However, in GR, it has been noted that the maximum mass and radius ($M$-$R$) of SS, using the conventional MIT bag model EoS, are $2.012~\mathrm{M_{\odot}}$ and 10.96 Km, respectively. Hence, the compactness is 0.27, which falls below the required range $(\frac{1}{3})$. 

To circumvent this difficulty, we introduce a newly proposed theory of gravity, {\it viz.}, $f(R+\alpha\,R^{2},T)$ \cite{Bhattacharjee} gravity, where $R$ and $T$, respectively, represent the Ricci scalar and trace of energy-momentum tensor, to study the GW echoes originating from SS. In this formalism, the parameter $\alpha$ quantifies the contribution from the quadratic curvature correction, while $\beta$ governs the strength of the non-minimal matter-gravity coupling. The combined effects may alter the hydrostatic equilibrium and internal structure of compact stars. Consequently, the modified TOV equations \cite{Tolman,Oppenheimer} can increase the maximum mass and compactness of stellar configurations, making GW echoes a potential probe of modified gravity effects. 

It is worth emphasising that the $f(R+\alpha R^{2})$ sector encompasses the well-known Starobinsky model \cite{Starobinsky}, which provides a successful inflationary scenario and remains in excellent agreement with the \textit{Planck} 2018 cosmological observations \cite{Akrami}. Consequently, $f(R+\alpha R^{2},T)$ theory constitutes a broader and more general extension of modified gravity, thereby extending and generalising the analyses presented in Refs.~\cite{Feola,Capozziello2}. We must highlight that motivated by precise astrophysical observations, modified gravity theories have gained significant attention as extensions of GR. Various approaches, including $f(R)$ \cite{Sotiriou,Felice,Nojiri}, $f(T)$ \cite{Cai}, $f(Q)$ \cite{Heisenberg}, $f(G)$ \cite{Nojiri1} and extended theories of gravity \cite{Capozziello1,Clifton,Bahamonde}, have been widely studied to address cosmological issues such as late-time cosmic acceleration, dark energy, and dark matter \cite{Sahni,Joyce,Peebles,Paddy}. In the context of stellar structures, these theories have also been applied to investigate and constrain their maximum mass limits \cite{Astashenok1,Astashenok7,Astashenok8}.

The present work is motivated by our recent investigation of SS in the quadratic curvature-matter coupled gravity $f(R,T)=R+\alpha R^{2}+2\beta T$ \cite{Bhattacharjee}. In that study, we derived the modified TOV equations and investigated the M-R properties and maximum-mass limits of MIT bag SS. We showed that the interplay between the quadratic curvature correction $\alpha$ and the non-minimal matter-gravity coupling $\beta$ can substantially modify the stellar compactness and permit maximum masses beyond the corresponding general-relativistic configurations, with masses up to $3.11\,\mathrm{M_{\odot}}$ for suitable choices of the model parameters. Importantly, that analysis was concerned primarily with the existence and equilibrium properties of massive strange-star configurations. The previous study raises a further question: whether the enhanced compactness produced by the same gravity-matter interaction can drive a physically stable strange star into the ultracompact regime required for GW echoes. The present work addresses this question. Rather than recalculating the maximum-mass problem, we use the modified stellar configurations as the starting point and investigate the additional geometric condition associated with the formation of a photon sphere and a GW trapping cavity. In particular, we determine the subset of the previously allowed parameter space for which the stellar radius becomes compatible with the photon-sphere condition, derive the corresponding echo producing mass-radius domain, and calculate the associated echo time delays and characteristic frequencies. 

The motivation for investigating GW echoes within the framework of $f(R+\alpha R^{2},T)$ gravity stems from the possibility that compact stars can serve as natural laboratories for testing strong-field gravity beyond the regime where GR has been extensively tested. In particular, SS, composed of ultra-dense deconfined quark matter beyond supra-saturation densities, may develop structural and space-time properties that differ appreciably from their GR counterparts when higher-order curvature corrections and gravity-matter coupling become significant. Such modifications can alter the effective potential governing gravitational perturbations and the associated photon-sphere trapping region, thereby influencing the formation and properties of GW echoes. While quasi-normal modes have traditionally been employed to probe compact object dynamics, GW echoes are expected to be more sensitive to the underlying space-time geometry and the presence of ultracompact configurations. Motivated by this, we investigate whether $f(R+\alpha R^{2},T)$ gravity can render SS sufficiently compact to produce detectable GW echoes. If confirmed, such echoes would provide a complementary observational probe for simultaneously constraining the EoS of ultra-dense matter and testing deviations from GR in the strong-field regime.

Our objective is not simply to demonstrate that modified gravity can produce more compact strange stars. We impose the photon-sphere condition together with the modified Buchdahl bound on the self-consistent TOV solutions. The intersection of the photon-sphere condition with the M-R sequences identifies the onset of the echo-producing configurations. We then obtain a restricted set of M-R configurations that satisfy both hydrostatic equilibrium and the compactness requirements for GW echoes. These configurations constitute an echo-specific M-R limit, which is distinct from the maximum mass obtained from the TOV sequence alone. This distinction is physically important. In GR, conventional MIT bag strange star configurations have a maximum compactness below the threshold M/R=1/3 required for a photon sphere. Our analysis shows that the combined quadratic-curvature and matter-geometry corrections can shift the stellar configurations into the photon-sphere regime. 

The paper is organised as follows: We obtain the modified TOV equations, within the framework of $f(R+\alpha\,R^{2},T)$ gravity, in Sec.~\ref{sec2}. In Sec.~\ref{sec3}, we discuss the GW echoes from SS: the conditions for echo emergence, the $M$-$R$ results and the typical echo properties. Sec.~\ref{sec3a} addresses the stability of the model, under radial perturbations, using Harrison-Zel'dovich-Novikov condition. In Sec.~\ref{sec3b}, we explore different observational signatures from the proposed framework and we summarise our main findings in Sec.~\ref{sec4}.  
\section{Tolman-Oppenheimer-Volkoff equations in $f(R+\alpha\,R^{2},T)$ gravity}\label{sec2}
We begin with the Einstein-Hilbert action in the form: 
\begin{equation}
	\mathcal{S}=\frac{1}{16\pi}\int{\sqrt{-g}f(\tilde{R},T)d^{4}x}+\int{\sqrt{-g}\mathcal{L_{M}}d^{4}x}, \label{eq1}
\end{equation}
Throughout this work, we adopt the geometrized system of units by setting $G=1$ and $c=1$. In Eq.~(\ref{eq1}), the modified curvature scalar is introduced as $\tilde{R}=R+\alpha R^{2}$, where $\alpha$ denotes the parameter characterizing the quadratic curvature correction. For the present study, we consider the functional form $f(\tilde{R},T)=R+\alpha R^{2}+2\beta T$, which incorporates both the quadratic Ricci scalar contribution and an explicit non-minimal coupling between geometry and matter through the trace of the energy-momentum tensor, $T$. The strength of the matter-geometry coupling is regulated by the parameter $\beta$. Here, $\mathcal{L}_{M}$ represents the matter Lagrangian density. Now, taking the variation of the action given in Eq.~(\ref{eq1}) with respect to the metric tensor $g_{ij}$ yields the modified gravitational field equations,
\begin{align}
	f_{R}R_{ij}-\frac{1}{2}g_{ij}f(\tilde{R},T)+\left(g_{ij}\Box-\nabla_{i}\nabla_{j}\right)f_{R}
	\nonumber\\\quad=8\pi T_{ij}-f_{T}T_{ij}-f_{T}\Theta_{ij},
	\label{eq2}
\end{align}
where $f_{R}=\partial f(\tilde{R},T)/\partial R$ and $f_{T}=\partial f(\tilde{R},T)/\partial T$ denote the partial derivatives of the gravitational Lagrangian with respect to the Ricci scalar and the trace of the energy-momentum tensor, respectively. The operator $\Box=\frac{1}{\sqrt{-g}}\partial_{i}\!\left(\sqrt{-g}\,g^{ij}\partial_{j}\right)$ represents the covariant d'Alembertian, while $\nabla_{i}$ is the covariant derivative associated with the Levi-Civita connection of the metric $g_{ij}$. Furthermore,
$\Theta_{ij}=g^{\alpha\beta}\frac{\delta\,T_{\alpha\beta}}{\delta g^{ij}}$, characterises the metric variation of the energy-momentum tensor. Substituting the specific functional form, $f(\tilde{R},T)=R+\alpha R^{2}+2\beta T$ into Eq.~(\ref{eq2}), the modified field equations reduce to
\begin{eqnarray}
	G_{ij}+2\alpha RR_{ij}-\frac{1}{2}\alpha R^{2}g_{ij}
	+2\alpha\left(g_{ij}\Box-\nabla_{i}\nabla_{j}\right)R
	\nonumber\\
	=8\pi T_{ij}-f_{T}T_{ij}-f_{T}\Theta_{ij}
	+\beta Tg_{ij},
	\label{eq3}
\end{eqnarray}
where $R_{ij}$ and $G_{ij}$ denote the Ricci and Einstein tensors, respectively. It is worth noting that the above formalism reproduces the $f(R,T)$ gravity proposed by Harko \textit{et al.}~\cite{Harko} in the limit $\alpha=0$ with $\beta\neq0$, whereas setting $\beta=0$ and retaining $\alpha\neq0$ recovers the quadratic $f(R)$ gravity model~\cite{Buchdahl1,Bertolami}.

To model the interior geometry of a static, spherically symmetric compact star, we consider the space-time metric:
\begin{equation}
	ds^{2}=-e^{2\nu(r)}dt^{2}+e^{2\lambda(r)}dr^{2}+r^{2}\left(d\theta^{2}+\sin^{2}\theta\,d\phi^{2}\right),
	\label{eq4}
\end{equation}
where the metric potentials $\nu(r)$ and $\lambda(r)$ are functions solely of the radial coordinate $r$.

Assuming the matter Lagrangian density to be $\mathcal{L}_{M}=p$, the matter content is described by the energy-momentum tensor of a perfect fluid,
\begin{equation}
	T_{ij}=(\rho+p)u_{i}u_{j}+pg_{ij},
	\label{eq5}
\end{equation}
where $\rho$ and $p$ denote the energy density and isotropic pressure, respectively, while the four-velocity satisfies the normalisation condition $u^{i}u_{i}=-1$. Under these assumptions, one obtains $T=-\rho+3p,\qquad \Theta_{ij}=-2T_{ij}+pg_{ij},\qquad\text{and}\qquad\Theta=-2T+4p$.

In the $f(\tilde{R},T)$ framework, the explicit dependence of the gravitational action on $T$ leads to a non-vanishing covariant divergence of the energy-momentum tensor. Consequently, the conservation equation assumes the form
\begin{align}
	\nabla^{i}T_{ij}
	=&\frac{f_{T}(\tilde{R},T)}
	{8\pi-f_{T}(\tilde{R},T)}
	\Big[
	(T_{ij}+\Theta_{ij})
	\nabla^{i}\ln f_{T}(\tilde{R},T)
	\nonumber\\&\quad
	+\nabla^{i}\Theta_{ij}
	-\frac{1}{2}g_{ij}\nabla^{i}T
	\Big].
	\label{eq6}
\end{align}

For the particular choice of model, the above expression simplifies considerably to the following form:
\begin{equation}
	\nabla^{i}T_{ij}=\frac{2\beta}{8\pi+2\beta}\left[\nabla^{i}
	\left(pg_{ij}-\frac{1}{2}g_{ij}\nabla^{i}T\right)\right].
	\label{eq7}
\end{equation}
This relation coincides with the corresponding non-conservation equation obtained by Pretel et al. \cite{Pretel} within the $f(R,T)$ gravity framework. Moreover, in the limiting case $\beta\rightarrow0$, the standard covariant conservation law of GR is recovered. 

Considering Eq.~(\ref{eq4}) together with the perfect fluid energy-momentum tensor in Eq.~(\ref{eq5}), the temporal and radial components of the modified field equations (\ref{eq3}) can be written, respectively, as
\begin{align}
	\frac{2\lambda'e^{-2\lambda}}{r}+\frac{1-e^{-2\lambda}}{r^{2}}
	+2\alpha R\Bigg[e^{-2\lambda}\Bigg\{-\nu''-\nu'^{2}
	+\lambda'\nu'-\nonumber\\\frac{2\nu'}{r}\Bigg\}\Bigg]
	+\frac{1}{2}\alpha R^{2}
	-2\alpha\Bigg[e^{-2\lambda}
	\Bigg\{R''+\left(\nu'-\lambda'+\frac{2}{r}\right)R'\Bigg\}\Bigg]\nonumber\\-2\alpha\,e^{-2\lambda}\nu'R'
	=8\pi\rho+3\beta\rho-\beta p,
	\label{eq8}
\end{align}
and
\begin{align}
	\frac{2\nu'e^{-2\lambda}}{r}
	-\frac{1-e^{-2\lambda}}{r^{2}}
	+2\alpha R\Bigg[e^{-2\lambda}
	\Bigg\{\nu''+\nu'^{2}-\lambda'\nu'
	\nonumber\\-\frac{2\lambda'}{r}\Bigg\}\Bigg]
	-\frac{1}{2}\alpha R^{2}
	+2\alpha\Bigg[e^{-2\lambda}
	\Bigg\{R''+\left(\nu'-\lambda'+\frac{2}{r}\right)R'\Bigg\}\Bigg]\nonumber\\-2\alpha\Big[R''-\lambda'R'\Big]
	=8\pi p+3\beta p-\beta\rho.
	\label{eq9}
\end{align}
Introducing the standard mass function through the relation, $e^{-2\lambda}=1-\frac{2m(r)}{r}$, and employing the adopted form of the gravitational Lagrangian, $f(\tilde{R},T)=R+\alpha R^{2}+2\beta T$, Eqs.~(\ref{eq7})-(\ref{eq9}) lead to the generalised Tolman-Oppenheimer-Volkoff (TOV) equations~\cite{Tolman,Oppenheimer}. The corresponding mass continuity equation is given by
\begin{align}
	\frac{dm}{dr}=&4\pi r^{2}\rho+\frac{\beta\,r^{2}}{2}\left(3\rho-p\right)+\frac{r^{2}}{2}\Bigg[2\alpha R\Bigg[e^{-2\lambda}\Bigg\{-\nu''\nonumber\\&\quad-\nu'^{2}+\lambda'\nu'
	-\frac{2\nu'}{r}\Bigg\}\Bigg]
	-\frac{1}{2}\alpha R^{2}
	+2\alpha
	\Bigg[e^{-2\lambda}
	\Bigg\{R''+\nonumber\\&\quad\left(\nu'-\lambda'
	+\frac{2}{r}\right)R'\Bigg\}\Bigg]+2\alpha\,e^{-2\lambda}\nu'R'
	\Bigg],
	\label{eq10}
\end{align}
while the hydrostatic equilibrium equation assumes the form
\begin{align}
	\frac{dp}{dr}
	=&\left(\frac{\beta}{8\pi+3\beta}\right)\rho'
	-\left(\frac{8\pi+2\beta}{8\pi+3\beta}\right)\times
	\nonumber\\&\quad\Bigg[(\rho+p)
	\left[
	\frac{m(r)+4\pi r^{3}p_{\rm eff}}
	{r\left[r-2m(r)\right]}
	\right]\Bigg],
	\label{eq11}
\end{align}
where the effective pressure is defined by
$p_{\rm eff}
=
p
+\frac{\beta}{8\pi}(3p-\rho)
-\frac{1}{8\pi}
\left[
2\alpha RR^{\,r}_{\,r}
-\frac{1}{2}\alpha R^{2}
+2\alpha\Box R
-2\alpha\nabla_{r}\nabla_{r}R
\right]$, and the prime denotes differentiation with respect to $r$.

The equilibrium structure of compact stars is determined by simultaneously integrating Eqs.~(\ref{eq10}) and (\ref{eq11}) for a prescribed EoS. The numerical integration is performed by imposing the regularity conditions at the stellar center, namely, $m(0)=0,~\rho(0)=\rho_{c}$, where $\rho_{c}$ represents the central energy density. The resulting solutions are subsequently employed to construct the corresponding $M$-$R$ relations. 
\section{GW echoes from strange stars}\label{sec3} 
The primary component in generating the GW echoes is the existence of a photon sphere. If a compact object contains a photon sphere, the signals from distant compact binary coalescences would fall on the surface and get reflected at the photon sphere. Following some time delay, multiple number of reflections and refractions may occur. Now, a photon sphere is obtained from the condition of unstable null geodesics. In the case of GR, the condition for null geodesics leads to $R_{\mathrm{ps}}=3M$, where $M$ is the total mass of the object. This is the standard radius of a photon sphere. Moreover, the maximum radius is obtained from Buchdahl bound, {\it viz.}, $R_{\mathrm{Buch}}=\frac{9M}{4}$ \cite{Buchdahl}. Hence, in GR, the radius of a compact object must lie in the range, $R_{\mathrm{Buch}}\leq\,R\leq\,R_{\mathrm{ps}}$ to generate echo. However, in modified gravity, the Buchdahl limit is modified, and consequently, the range of radius is modified. It has been shown that in case of modified gravity, the Buchdahl bound is revised in the following form, $\frac{M}{R}=\frac{4}{9}-\frac{a}{6}$ \cite{Burikham}, where $a=4\pi\,p_{eff}(R)R^{2}$ and $p_{eff}$ has been expressed earlier. Further, in this theory of gravity, we have noted that in the vacuum and asymptotic limit, the exterior solution resembles the Schwarzschild one. Hence, the condition for $R_{\mathrm{ps}}=3M$ remains valid here. Hence, in the present context, the compactness must lie in the range, $\frac{1}{3}\leq\frac{M}{R}\leq\left(\frac{4}{9}-\frac{a}{6}\right)$ to generate GW echoes.     

The typical light time from the stellar centre to the photon sphere is termed the echo time, and it is expressed as:
\begin{equation}
	\tau_{\text{echo}}=\bigintssss_{0}^{3M}\frac{dr}{\sqrt{e^{2\nu(r)}\left(1-\frac{2M(r)}{r}\right)}}, \label{eq12}
\end{equation}
where, $\nu(r)$ and $M(r)$ are determined from the solution of TOV equations. The corresponding echo frequency can be computed as \cite{Cardoso,Cardoso2,Mark}:
\begin{equation}
	f_{\text{echo}}=\frac{\pi}{\tau_{\text{echo}}}. \label{eq13}
\end{equation}
Previously, in the work of Abedi and Afshordi \cite{Abedi}, the echo frequency has been estimated as, $f_{\text{echo}}=\frac{1}{2\tau_{\text{echo}}}$. This $f_{\text{echo}}$ must signify the repetition frequency of the GW signal. The reason we are choosing Eq.~(\ref{eq13}) is that the echoes and the corresponding frequencies are associated with standing waves inside the photon sphere. During the merger, these modes are excited and partly trapped inside the photon sphere \cite{Kokkotas,Andersson}. After some time delay, these signals are transmitted with approximately the same frequency as that of the standing waves. Hence, the GW echo frequency is set by the eigen-modes of the photon sphere trapping cavity. Consequently, it is governed by the geometry and boundary conditions of the trapping region and is generally independent of the GW frequency emitted during the inspiral stage of the merger.

To study the GW echoes from strange stars, we consider the following:
\begin{itemize}
	\item The interior strange matter distribution is described by the MIT bag model EoS of the form \cite{Kettner}: $p=\frac{1}{3}(\rho-4B_{g})$, where $B_{g}$ is the bag constant. 
	\item Following the work of Madsen \cite{Madsen}, $B_{g}$ is varied within the stable strange matter range, i.e., $57.55\leq\,B_{g}\leq\,95.11$ \\$\mathrm{MeV/fm^{3}}$. 
	\item Suitable combinations of the non-minimal matter coupling parameter, $\beta$ and the quadratic curvature corrections, $\alpha$ are considered based on the physically acceptable TOV solutions. 
\end{itemize} 

\begin{figure*}[t!]
	\begin{minipage}{0.4\textwidth}
		\centering
		\includegraphics[width=8cm]{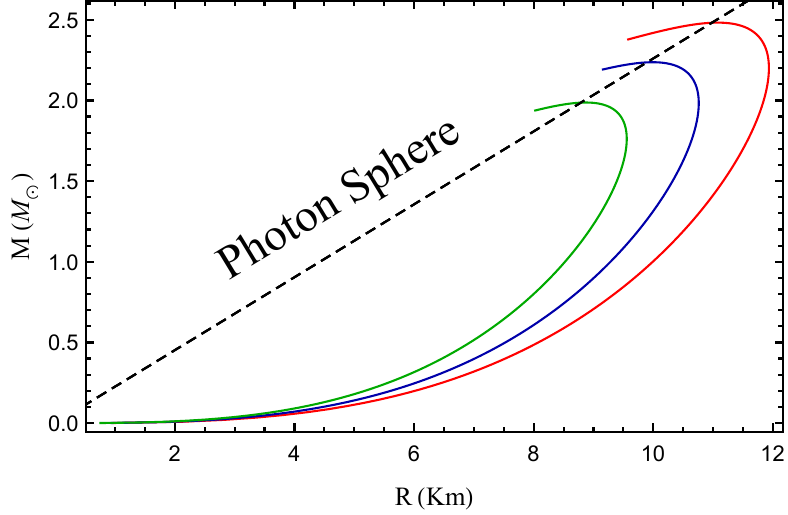}
		\subcaption{}
		\label{fig1}
	\end{minipage}
	\hfil
	\begin{minipage}{0.4\textwidth}
		\centering
		\includegraphics[width=8cm]{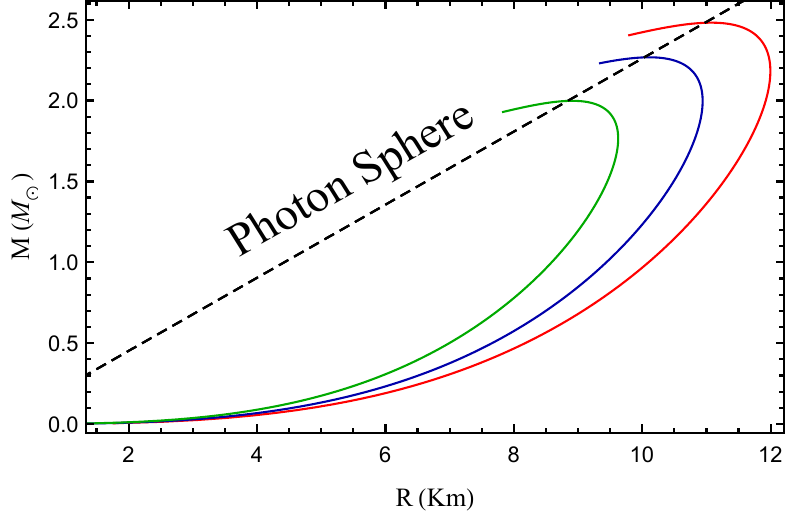}
		\subcaption{}
		\label{fig2}
	\end{minipage}
	\caption{\raggedright $M$-$R$ plots. Here, panel (a) corresponds to $\beta=-0.5$, with $\alpha=10,\, 8.1,$ and $6.4$ represented by the red, blue, and green lines, respectively; and panel (b) $\alpha=5.0$, with $\beta=-0.98,\,-0.83,$ and $-0.65$ represented by the red, blue, and green lines, respectively.}
\end{figure*}

In line with the above, the solutions of TOV equations, resulting in $M$-$R$ plots, are shown in Figs.~\ref{fig1} and \ref{fig2}. 
\begin{table*}
	\caption{\raggedright Tabulation of $M$-$R$ values, corresponding to (a) Fig.~\ref{fig1}, and (b) Fig.~\ref{fig2}, under suitable parametric choices of $\alpha$ and $\beta$. Here, panel (a) corresponds to $\beta=-0.5$ and $\alpha=10,~8.1 ~\text{and},~6.4$, respectively and panel (b) corresponds to $\alpha=5.0$ and $\beta=-0.98,~-0.83 ~\text{and},~-0.65$, respectively.}
	\begin{minipage}{0.4\textwidth}
		\subcaption{}
		\label{tab1}
		\begin{tabular}{cccccc}
			\hline
			\hline
			\multirow{2}{*}{$\beta$} & \multirow{2}{*}{$\alpha$} & $B_{g}$ & $M$ & $R$ & \multirow{2}{*}{$\frac{M}{R}$} \\ 
			&& $(\mathrm{MeV/fm^{3}})$ & $(\mathrm{M_{\odot}})$ & (Km) & \\
			\hline
			\multirow{3}{*}{-0.5} & 10 & 61 & 2.482 & 11.07 & 0.330488\\
			& 8.1 & 75 & 2.23659 & 9.99 & 0.330237 \\
			& 6.4 & 95.11 & 1.9869 & 8.87 & 0.3303 \\
			\hline
			\hline
		\end{tabular}
	\end{minipage}
	\hfil
	\begin{minipage}{0.4\textwidth}
		\subcaption{}
		\label{tab2}
		\begin{tabular}{cccccc}
			\hline
			\hline
			\multirow{2}{*}{$\alpha$} & \multirow{2}{*}{$\beta$} & $B_{g}$ & $M$ & $R$ & \multirow{2}{*}{$\frac{M}{R}$} \\ 
			&& $(\mathrm{MeV/fm^{3}})$ & $(\mathrm{M_{\odot}})$ & (Km) & \\
			\hline
			\multirow{3}{*}{5} & -0.98 & 63.40 & 2.48146 & 11.0751 & 0.330484 \\
			& -0.83 & 75 & 2.26738 & 10.1096 & 0.330811 \\
			& -0.65 & 95.11 & 1.99824 & 8.91077 & 0.330768\\
			\hline
			\hline
		\end{tabular}
	\end{minipage}
\end{table*}

Figs.~\ref{fig1} and \ref{fig2}, together with Tables~\ref{tab1} and \ref{tab2} describe the maximum permissible mass and radius values from the hydrostatic equilibrium of TOV equations. We observe that the use of $f(R+\alpha\,R^{2},T)$ gravity effectively increases the compactness of SS beyond their general relativistic counterpart. From these results, we note the following features: (i) for a given value of $\beta$, the corresponding range of $\alpha$, and vice versa, is determined up to the limit at which the TOV equations yield physically realistic solutions, (ii) to ensure a ghost-field free scenario, we have invoked the positive kinetic term by imposing the condition, $f_{RR}\geq0$, implying, $\alpha>0$, (iii) we have noted that an increased compactness, beyond 0.33, is only obtained for $\beta<0$. Accordingly, we have considered the range of $\beta$, (iv) although, for non-interacting stable strange matter, $B_{g}$ varies within the range $57.55\leq\,B_{g}\leq95.11~\mathrm{MeV/fm^{3}}$ \cite{Madsen}, we find that to investigate GW echoes while preserving the presence of photon sphere within the allowed parameter space, the lower bound of $B_{g}$ must be increased to $61~\mathrm{MeV/fm^{3}}$ for $\beta=-0.5$ and $\alpha=10$, (v) with increasing $B_{g}$, the pressure decreases and the EoS becomes softer. Consequently, the $M$-$R$ values decrease, which is evident from Tables~\ref{tab1} and \ref{tab2}. 

The intersection of the photon sphere with $M$-$R$ curves reveal a very interesting result: the nodes of intersection are the brinks of GW echoe producing configurations. Below these points, photon sphere is absent and echo is not possible. Now, from the solution of TOV equations, we have tabulated the $M$-$R$ values and the corresponding compactnesses in Tables~\ref{tab1} and \ref{tab2}. However, filtering the results of the TOV solutions by imposing the modified Buchdahl limit \cite{Burikham}, together with the compactness limit $(\frac{1}{3})$, we note that the corresponding $M$-$R$ values and compactnesses are modified. We tabulate the results in Tables~\ref{tab3} and \ref{tab4}. 

\begin{table*}[t!]
	\caption{\raggedright New $M$-$R$ limits of SS from the aspect of photon sphere and GW echo. Each row corresponds to the row of (a) Table~\ref{tab1}, and (b) Table~\ref{tab2}. Here, panel (a) corresponds to $\beta=-0.5$ and $\alpha=10,~8.1 ~\text{and},~6.4$, respectively and panel (b) corresponds to $\alpha=5.0$ and $\beta=-0.98,~-0.83 ~\text{and},~-0.65$, respectively.}
	\begin{minipage}{0.4\textwidth}
		\centering
		\subcaption{}
		\label{tab3}
		\begin{tabular}{cccc}
			\hline
			\hline
			$M$ & $R$ & \multirow{2}{*}{a} & \multirow{2}{*}{$\frac{M}{R}$}\\
			($\mathrm{M_{\odot}}$) & (Km) & & \\ 
			\hline
			2.481 & 10.98 & -0.0652 & 0.333619 \\
			2.235 & 9.89 & -0.0649 & 0.333395 \\
			1.98614 & 8.79 & -0.0650 & 0.333318 \\ \hline
			\hline				 
		\end{tabular}
	\end{minipage}
	\hfil
	\begin{minipage}{0.4\textwidth}
		\centering
		\subcaption{}
		\label{tab4}
		\begin{tabular}{cccc}
			\hline
			\hline
			$M$ & $R$ & \multirow{2}{*}{a} & \multirow{2}{*}{$\frac{M}{R}$}\\
			($\mathrm{M_{\odot}}$) & (Km) & & \\
			\hline
			2.48068 & 10.977 &	-0.0485 & 0.333395 \\
			2.26682 & 10.0307 & -0.0532 & 0.333356 \\
			1.9977 & 8.83983 & -0.0593 & 0.333365 \\
			\hline
			\hline
		\end{tabular}
	\end{minipage}
\end{table*}

\begin{figure*}[t!]
	\begin{minipage}{0.4\textwidth}
		\centering
		\includegraphics[width=8cm]{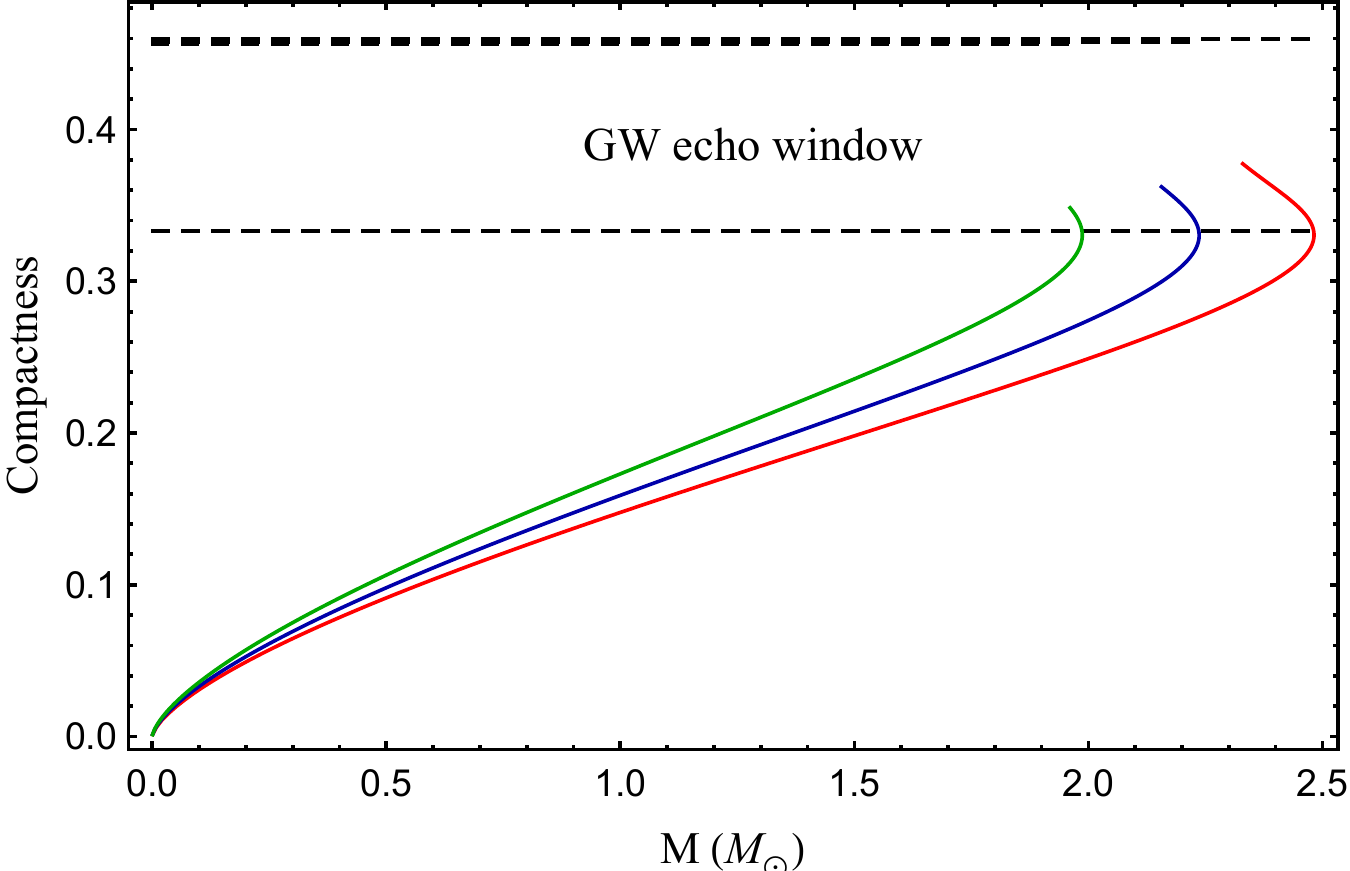}
		\subcaption{}
		\label{fig3}
	\end{minipage}
	\hfil
	\begin{minipage}{0.4\textwidth}
		\centering
		\includegraphics[width=8cm]{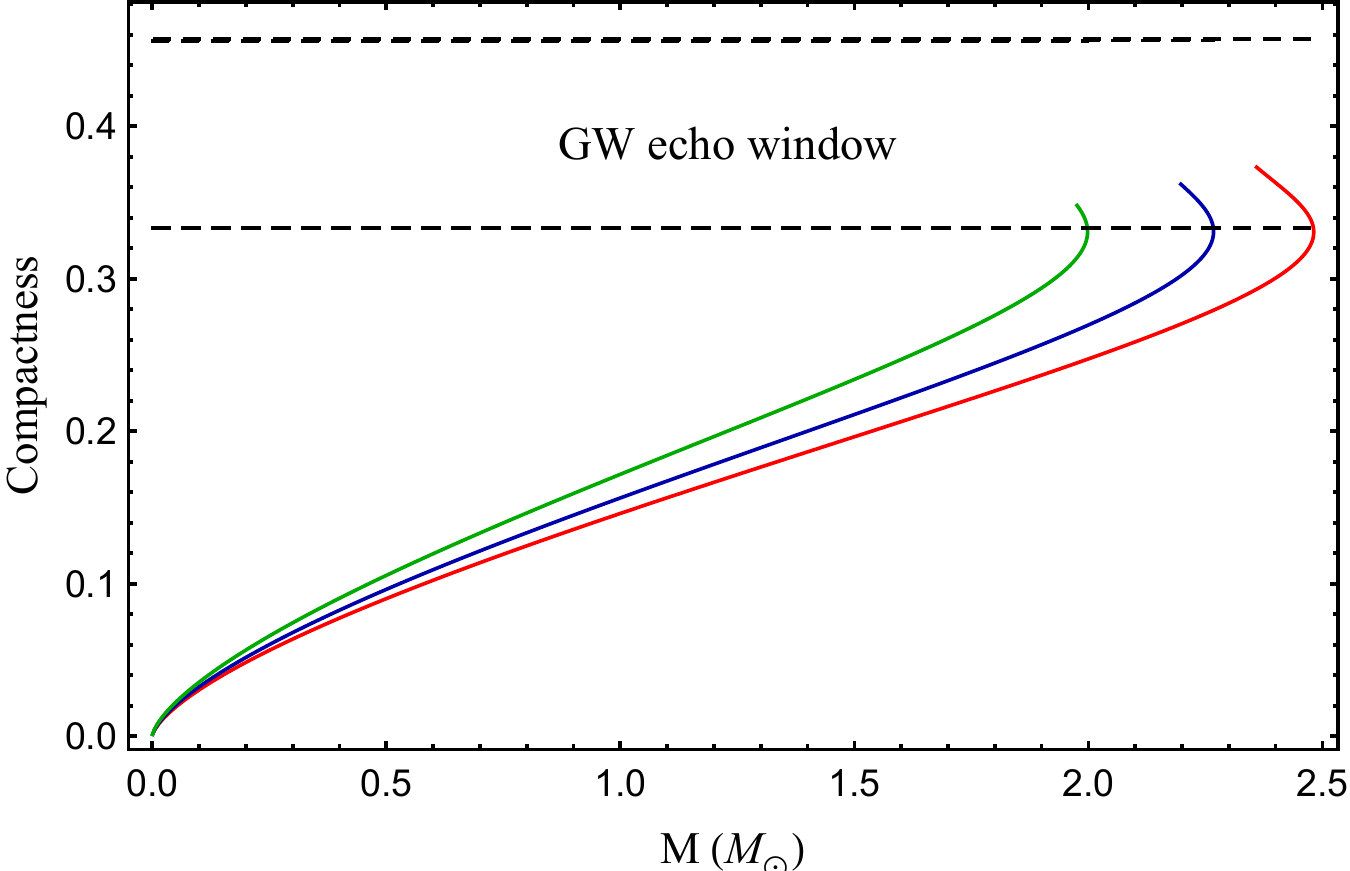}
		\subcaption{}
		\label{fig4}
	\end{minipage}
	\caption{\raggedright Variation of compactness with maximum mass $(\mathrm{M_{\odot}})$ for the newly obtained mass limit of SS in the context of GW echoes. Here, panel (a) corresponds to $\beta=-0.5$, with $\alpha=10,\, 8.1,$ and $6.4$ represented by the red, blue, and green lines, respectively; and panel (b) $\alpha=5.0$, with $\beta=-0.98,\,-0.83,$ and $-0.65$ represented by the red, blue, and green lines, respectively.}
\end{figure*}

It is evident that the $M$-$R$ values are lower than those tabulated in Tables~\ref{tab1} and \ref{tab2}. Hence, from consideration of hydrostatic equilibrium, the $M$-$R$ results obtained in Tables~\ref{tab1} and \ref{tab2} remains valid for SS, however, to investigate GW echoes, the allowed $M$-$R$ values must correspond to those tabulated in Tables~\ref{tab3} and \ref{tab4}. Within the present parameter space and considering the aspects of GW echoes, the $M$-$R$ values of Tables~\ref{tab3} and \ref{tab4} may be termed the new $M$-$R$ limits of SS. Moreover, within the newly obtained mass limit, we present the variation of the compactness with the maximum mass in Figs.~\ref{fig3} and \ref{fig4}. The figures also indicate the compactness window conducive to the generation of GW echoes, as inferred from the modified Buchdahl bound \cite{Burikham}.

From Figs.~\ref{fig3} and \ref{fig4}, it must be noted that, within the compactness window, these mass points can be readily employed to study GW echoes. Further, Burikham et al. \cite{Burikham} reported that the parameter `a' appearing in the modified Buchdahl limit is negative. In the present model, we have shown the variation of `a' with maximum mass, to substantiate the finding. 
\begin{figure*}[t!]
	\begin{minipage}{0.4\textwidth}
		\centering
		\includegraphics[width=8cm]{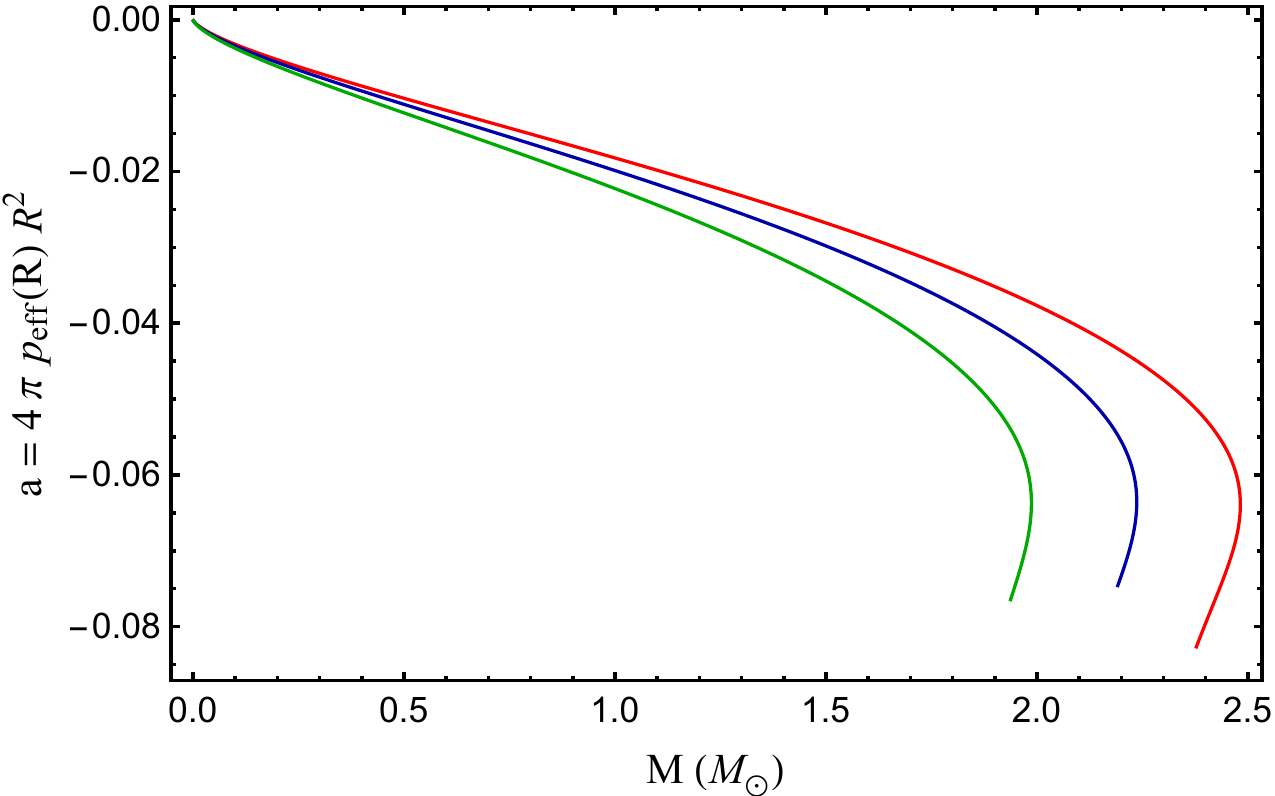}
		\subcaption{}
		\label{fig5}
	\end{minipage}
	\hfil
	\begin{minipage}{0.4\textwidth}
		\centering
		\includegraphics[width=8cm]{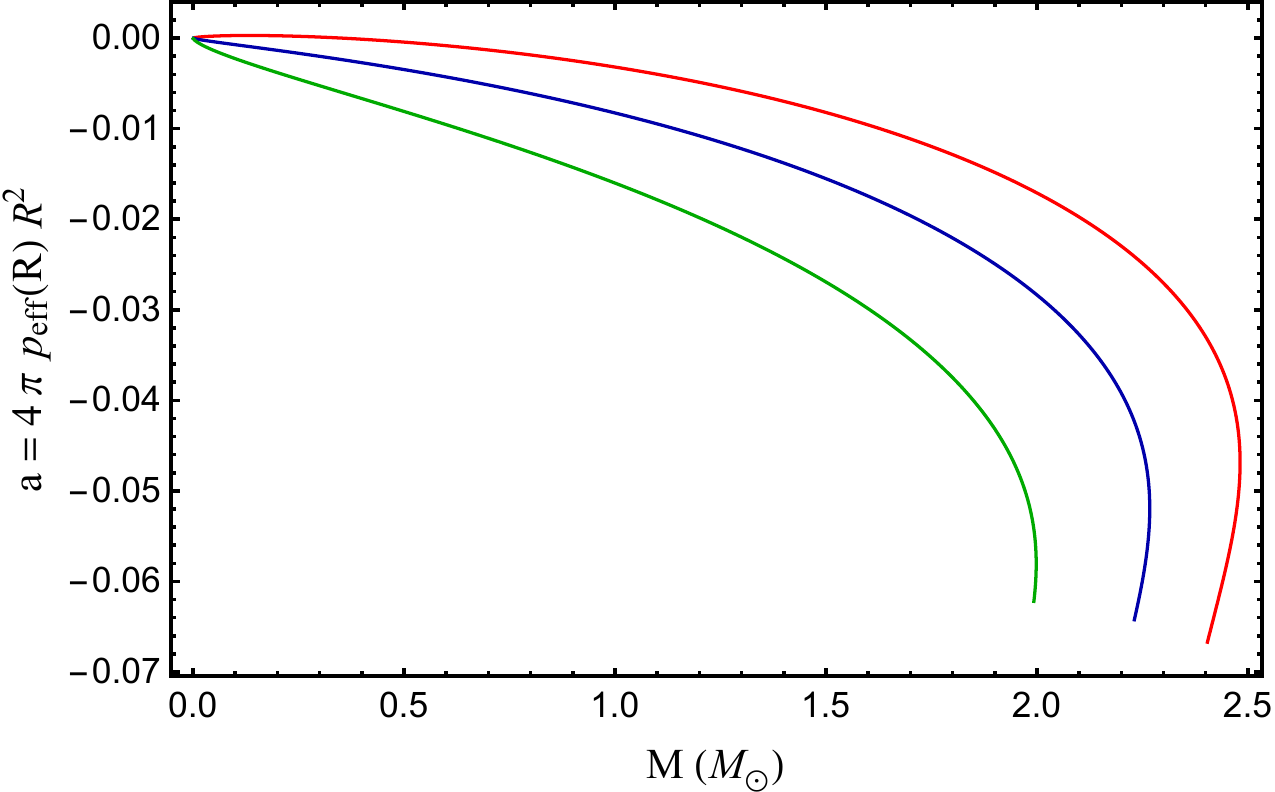}
		\subcaption{}
		\label{fig6}
	\end{minipage}
	\caption{\raggedright Variation of `a' with maximum mass $(\mathrm{M_{\odot}})$. Here, panel (a) corresponds to $\beta=-0.5$, with $\alpha=10,\, 8.1,$ and $6.4$ represented by the red, blue, and green lines, respectively; and panel (b) $\alpha=5.0$, with $\beta=-0.98,\,-0.83,$ and $-0.65$ represented by the red, blue, and green lines, respectively.}
\end{figure*}

\begin{figure*}[t!]
	\begin{minipage}{0.4\textwidth}
		\centering
		\includegraphics[width=8cm]{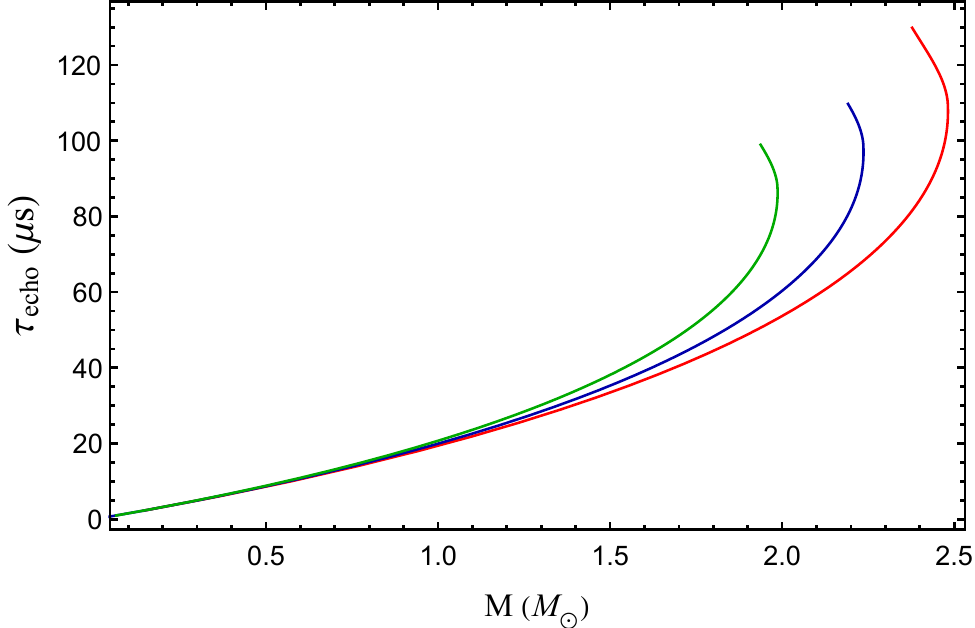}
		\subcaption{}
		\label{fig7}
	\end{minipage}
	\hfil
	\begin{minipage}{0.4\textwidth}
		\centering
		\includegraphics[width=8cm]{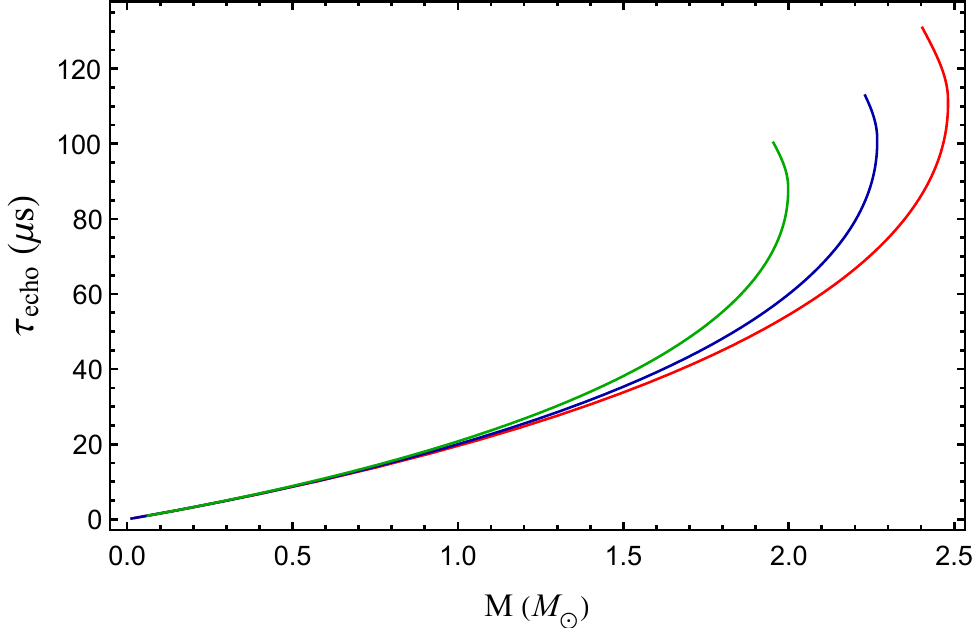}
		\subcaption{}
		\label{fig8}
	\end{minipage}
	\caption{\raggedright Variation of echo time $(\tau_{\text{echo}})$ with maximum mass $(\mathrm{M_{\odot}})$. Here, panel (a) corresponds to $\beta=-0.5$, with $\alpha=10,\, 8.1,$ and $6.4$ represented by the red, blue, and green lines, respectively; and panel (b) $\alpha=5.0$, with $\beta=-0.98,\,-0.83,$ and $-0.65$ represented by the red, blue, and green lines, respectively.}
\end{figure*}

Since, we have established the fact that the new mass limits can be used to study echoes, we proceed to studying the properties of GW echoes. In this present context, the parametric variations of $\alpha$, $\beta$ and $B_{g}$ produces different $M$-$R$-compactness results. Consequently, the characteristics of GW echoes from SS in this framework will be different for each combination. Following Eq.~(\ref{eq12}), we show the variation of echo time with maximum mass and the results are illustrated in Figs.~\ref{fig7} and \ref{fig8}.

\begin{figure*}[t!]
	\begin{minipage}{0.4\textwidth}
		\centering
		\includegraphics[width=8cm]{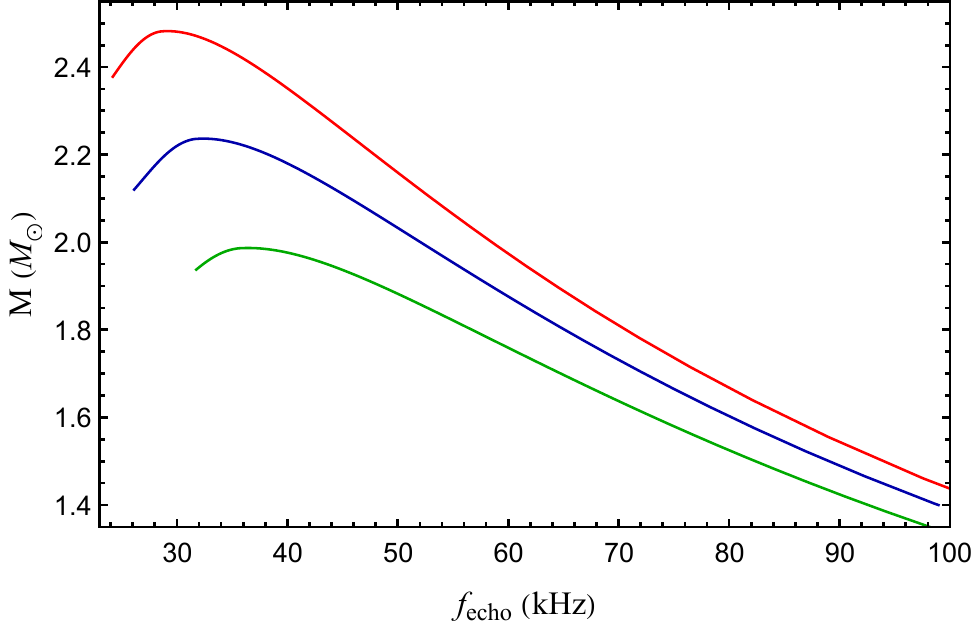}
		\subcaption{}
		\label{fig9}
	\end{minipage}
	\hfil
	\begin{minipage}{0.4\textwidth}
		\centering
		\includegraphics[width=8cm]{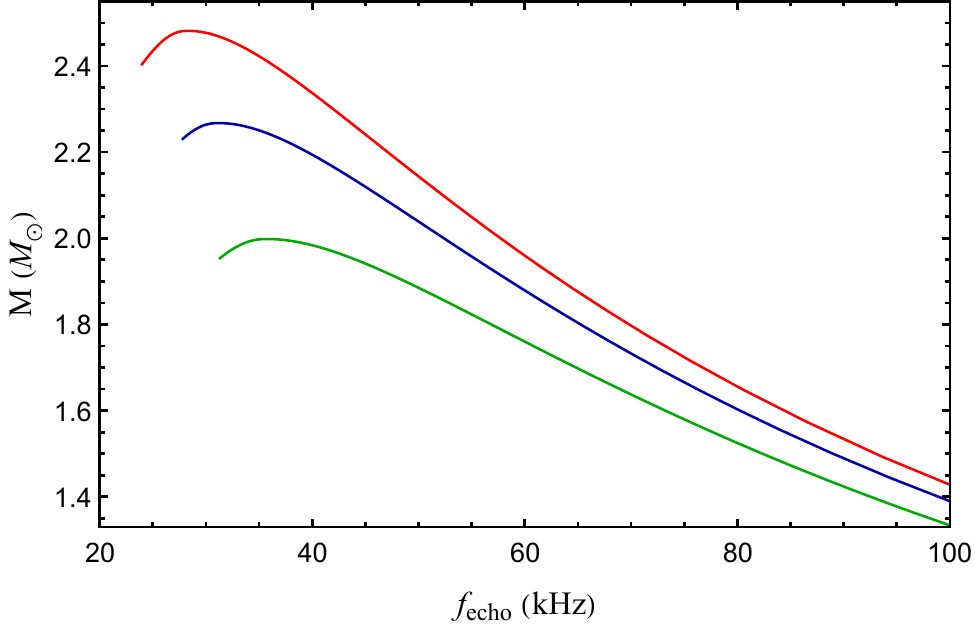}
		\subcaption{}
		\label{fig10}
	\end{minipage}
	\caption{\raggedright Variation of maximum mass $(\mathrm{M_{\odot}})$ with echo frequency $(f_{\text{echo}})$. The peaks describe the newly obtained maximum mass limit of SS in the context of GW echoes. Here, panel (a) corresponds to $\beta=-0.5$, with $\alpha=10,\, 8.1,$ and $6.4$ represented by the red, blue, and green lines, respectively; and panel (b) $\alpha=5.0$, with $\beta=-0.98,\,-0.83,$ and $-0.65$ represented by the red, blue, and green lines, respectively.}
\end{figure*}

From Figs.~\ref{fig7} and \ref{fig8}, it is noted that with increasing $B_{g}$, the echo time decreases. However, this decreasing $\tau_{\text{echo}}$ is dependent on the parametric combinations of $\alpha$, $\beta$ and $B_{g}$, as described in Tables~\ref{tab1} and \ref{tab2}. For each combination with increasing $B_{g}$, the pressure decreases in accordance with MIT bag model and EoS becomes softer. A softer EoS leads to a more compressed stellar configuration. Hence, the radius decreases, and an increased gravitational pull may shift the position of photon sphere and the potential barrier. Consequently, the GW signal travels a reduced effective cavity length, and the echo time decreases. As a corollary, with decreasing $\tau_{\text{echo}}$, the echo frequency increases, as demonstrated in Figs.~\ref{fig9} and \ref{fig10}. 
From Figs.~\ref{fig7}-\ref{fig10}, we extract echo times and frequencies for the newly obtained maximum mass limit of SS and are tabulated in Tables~\ref{tab5} and \ref{tab6}. 
\begin{table*}[t!]
	\caption{\raggedright Tabulation of echo parameters within the allowed parameter space. Here, panel (a) corresponds to $\beta=-0.5$ and $\alpha=10,~8.1 ~\text{and},~6.4$, respectively and panel (b) corresponds to $\alpha=5.0$ and $\beta=-0.98,~-0.83 ~\text{and} -0.65$, respectively.}
	\begin{minipage}{0.45\textwidth}
		\centering
		\subcaption{}
		\label{tab5}
		\begin{tabular}{ccccc}
			\hline
			\hline
			$\beta$ & $\alpha$ & $B_{g}~(\mathrm{MeV/fm^{3}})$ & $\tau_{\text{echo}}~(\mathrm{\mu\,s})$ & $f_{\text{echo}}~(\mathrm{kHz})$ \\
			\hline
			\multirow{3}{*}{-0.5} & 10 & 61 & 107.40 & 29.25 \\
			& 8.1 & 75 & 96.86 & 32.43 \\
			& 6.4 & 95.11 & 86.16 & 36.46 \\ 
			\hline
			\hline
		\end{tabular}
	\end{minipage}
	\hfil
	\begin{minipage}{0.45\textwidth}
		\subcaption{}
		\label{tab6}
		\begin{tabular}{ccccc}
			\hline
			\hline
			$\alpha$ & $\beta$ & $B_{g}~(\mathrm{MeV/fm^{3}})$ & $\tau_{\text{echo}}~(\mathrm{\mu\,s})$ & $f_{\text{echo}}~(\mathrm{kHz})$\\
			\hline
			\multirow{3}{*}{5} & -0.98 & 63.40 & 110.46 & 28.44 \\
			& -0.83 & 75 & 100.28 & 31.33 \\
			& -0.65 & 95.11 & 87.67 & 35.83 \\
			\hline
			\hline
		\end{tabular}
	\end{minipage}
\end{table*}

We observe that the obtained echo frequencies are in the kHz range which is particularly interesting since the kHz frequency band is very sensitive to the post-merger signals of binary NS, occurring at an estimated rate of approximately 0.1 $\mathrm{year^{-1}}$. Hence, the efficacy of the present framework is three-folds: (i) it can be implemented to study the post-merger signals of GW echoes from compact binary coalescences, (ii) the detection of echo signals within this frequency band could provide evidence that originate from SS stars in quadratic curvature-matter coupled gravity, and (iii) in this framework, the existence of photon sphere and the associated emission of GW echoes place stringent bounds on the maximum mass of SS. Consequently, these limits define a new maximum mass of SS from the perspective of GW echoes. 
\section{Stability under Harrison-Zel'dovich-Novikov condition}\label{sec3a} 
\begin{figure*}[t!]
	\begin{minipage}{0.4\textwidth}
		\centering
		\includegraphics[width=8cm]{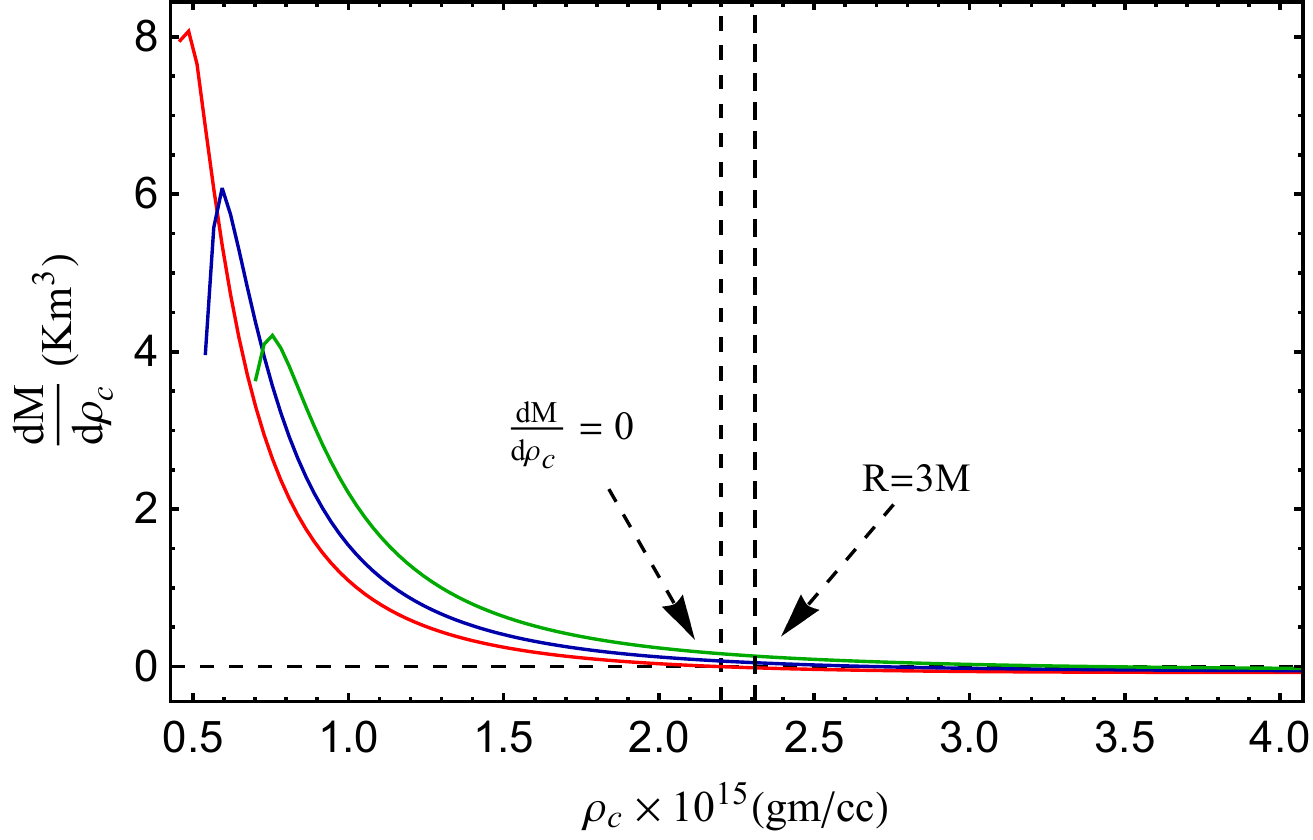}
		\subcaption{}
		\label{fig11}
	\end{minipage}
	\hfil
	\begin{minipage}{0.4\textwidth}
		\centering
		\includegraphics[width=8cm]{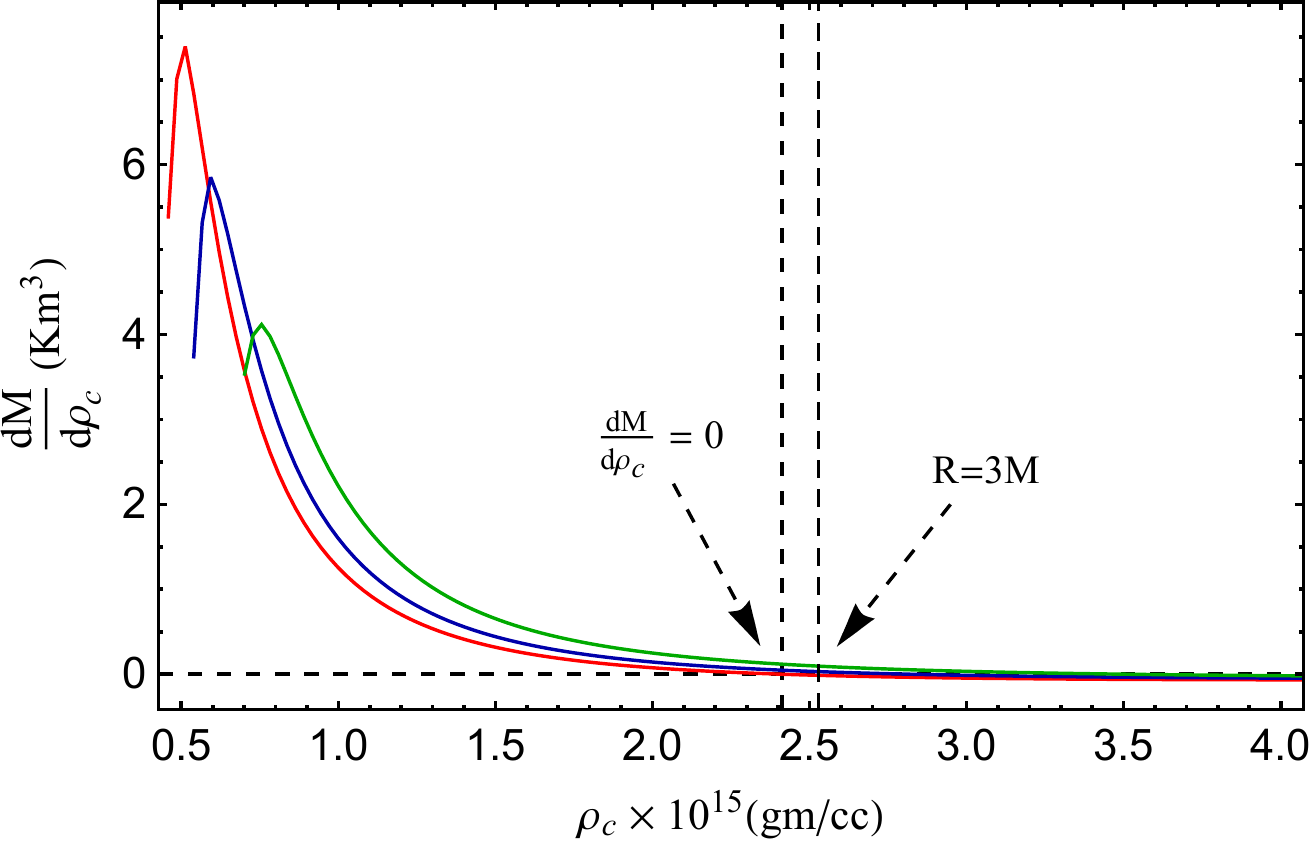}
		\subcaption{}
		\label{fig12}
	\end{minipage}
	\caption{\raggedright Mass vs central density plot. Here, panel (a) corresponds to $\beta=-0.5$, with $\alpha=10,\, 8.1,$ and $6.4$ represented by the red, blue, and green lines, respectively; and panel (b) $\alpha=5.0$, with $\beta=-0.98,\,-0.83,$ and $-0.65$ represented by the red, blue, and green lines, respectively.}
\end{figure*}

The Harrison-Zel'dovich-Novikov (HZN) condition \cite{Zeldovich} is widely employed to investigate the stability of equilibrium stellar configurations. This criterion dictates the onset where the compact star transitions from a stable to an unstable configuration under radial perturbations. Consequently, this condition directly governs the real oscillation frequency of the fundamental radial mode (f-mode). The criterion is commonly examined through the variation of the gravitational mass, $M$, with respect to the central density, $\rho_{c}$, particularly in the vicinity of the maximum-mass configuration. By solving the TOV equations, one obtains a sequence of equilibrium configurations for progressively increasing values of $\rho_{c}$, eventually reaching a limiting configuration corresponding to the maximum mass. Beyond this point, a further increase in the central density does not lead to a stable increase in the stellar mass; instead, the configuration enters an unstable regime in which gravitational effects dominate the pressure support, ultimately resulting in gravitational collapse. As we have shown that GW echoes constrain the maximum mass of strange stars, it is inevitable to study the stability of these obtained maximum mass limits within the present parameter space. This aspect has been graphically illustrated in Figs.~\ref{fig11} and \ref{fig12}. 

In Figs.~\ref{fig11} and \ref{fig12}, we have shown the standard HZN condition, along with the stability of the newly obtained mass limit by imposing the photon sphere constraint. We note that the maximum masses, as tabulated in Tables~\ref{tab1}-\ref{tab4}, maintain the condition $\frac{dM}{d\rho_{c}}\approx0$ indicating stellar structures in stable equilibrium. 

\section{Observational signatures and model constraints}\label{sec3b}
\subsection{Optical shadows and EHT constraints} Beyond post-merger GW echoes, the presence of photon sphere in these ultracompact strange star configurations provides a direct observational link to electromagnetic imaging, such as that conducted by the Event Horizon Telescope (EHT) \cite{EHT}. In the present formalism of a static, spherically symmetric space-time, the exterior space-time matches the Schwarzschild metric outside the stellar radius. Consequently, the apparent angular diameter $\theta_{\text{sh}}$ of the photon ring/shadow as seen by a distant observer at distance $D$ is given geometrically by the impact parameter of light rays at the photon sphere:
\begin{equation}
	b_{\text{ps}}=\frac{R_{\text{ps}}}{\sqrt{e^{2\nu(R_{\text{ps}})}}}. \label{eq14}
\end{equation}

\begin{figure*}[t!]
	\begin{minipage}{0.4\textwidth}
		\centering
		\includegraphics[width=8cm]{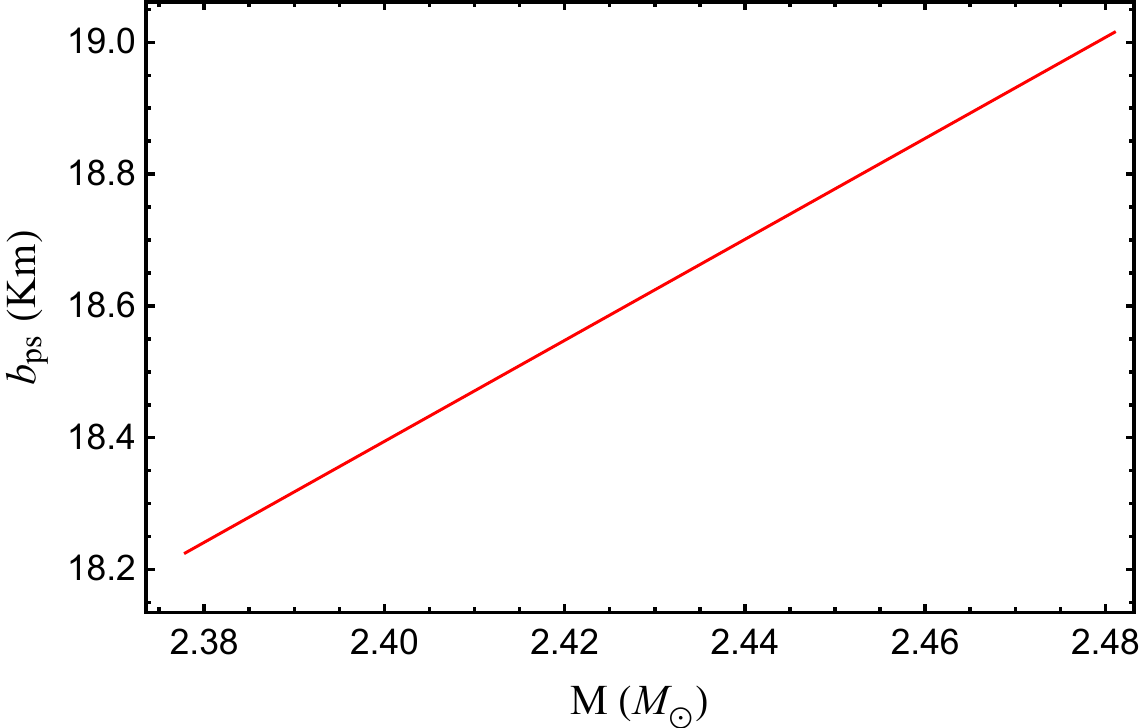}
		\subcaption{}
		\label{fig13}
	\end{minipage}
	\hfil
	\begin{minipage}{0.4\textwidth}
		\centering
		\includegraphics[width=8cm]{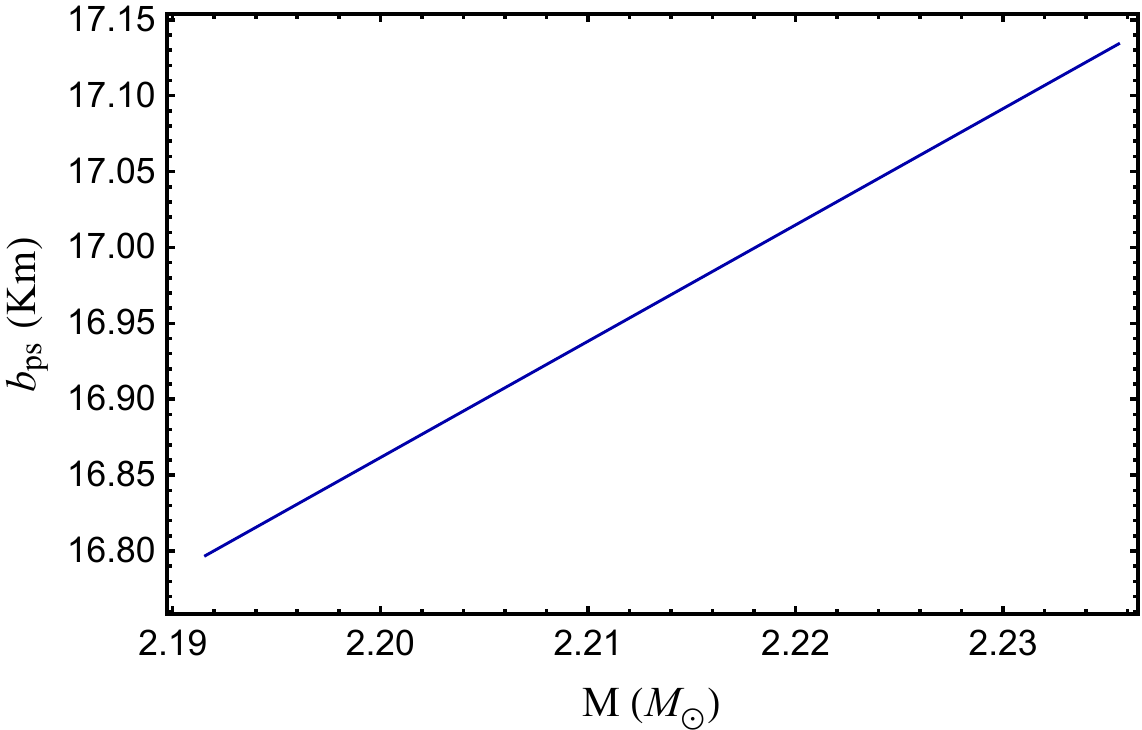}
		\subcaption{}
		\label{fig14}
	\end{minipage}
	\hfil
	\begin{minipage}{0.4\textwidth}
		\centering
		\includegraphics[width=8cm]{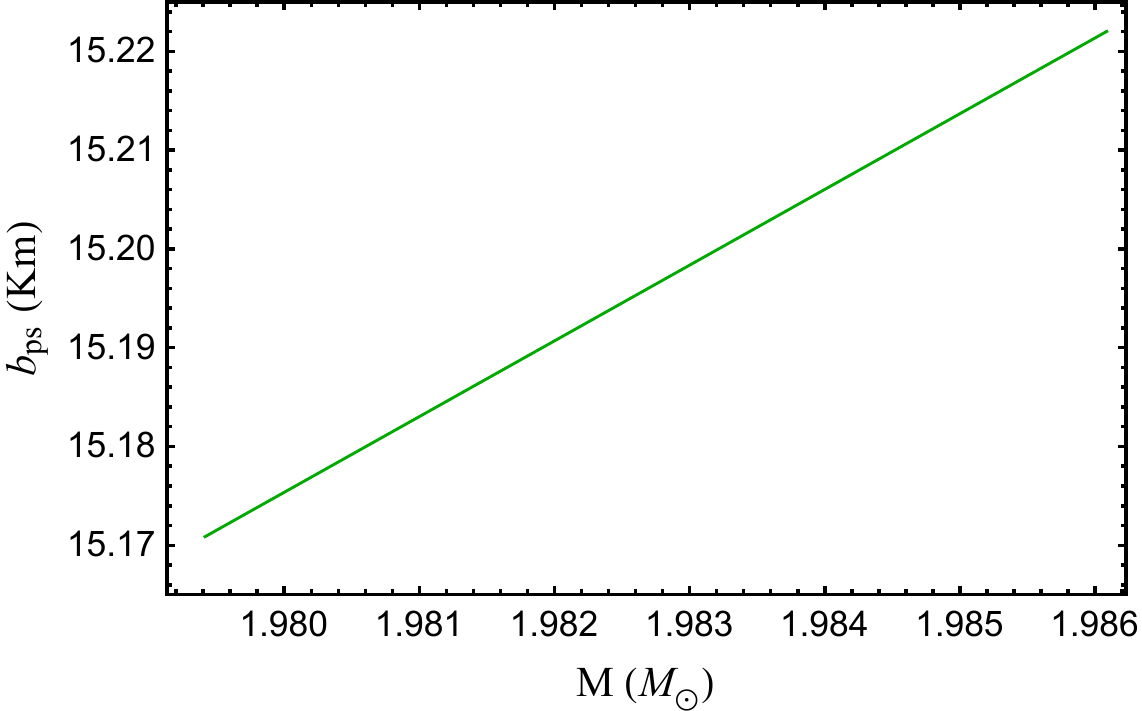}
		\subcaption{}
		\label{fig15}
	\end{minipage}
	\caption{\raggedright Shadow impact parameter vs mass plot for (a) $\beta=-0.5$ and $\alpha=10$, (b) $\beta=-0.5$ and $\alpha=8.1$, and (c) $\beta=-0.5$ and $\alpha=6.4$.}
\end{figure*}

\begin{figure*}
	\begin{minipage}{0.4\textwidth}
		\centering
		\includegraphics[width=8cm]{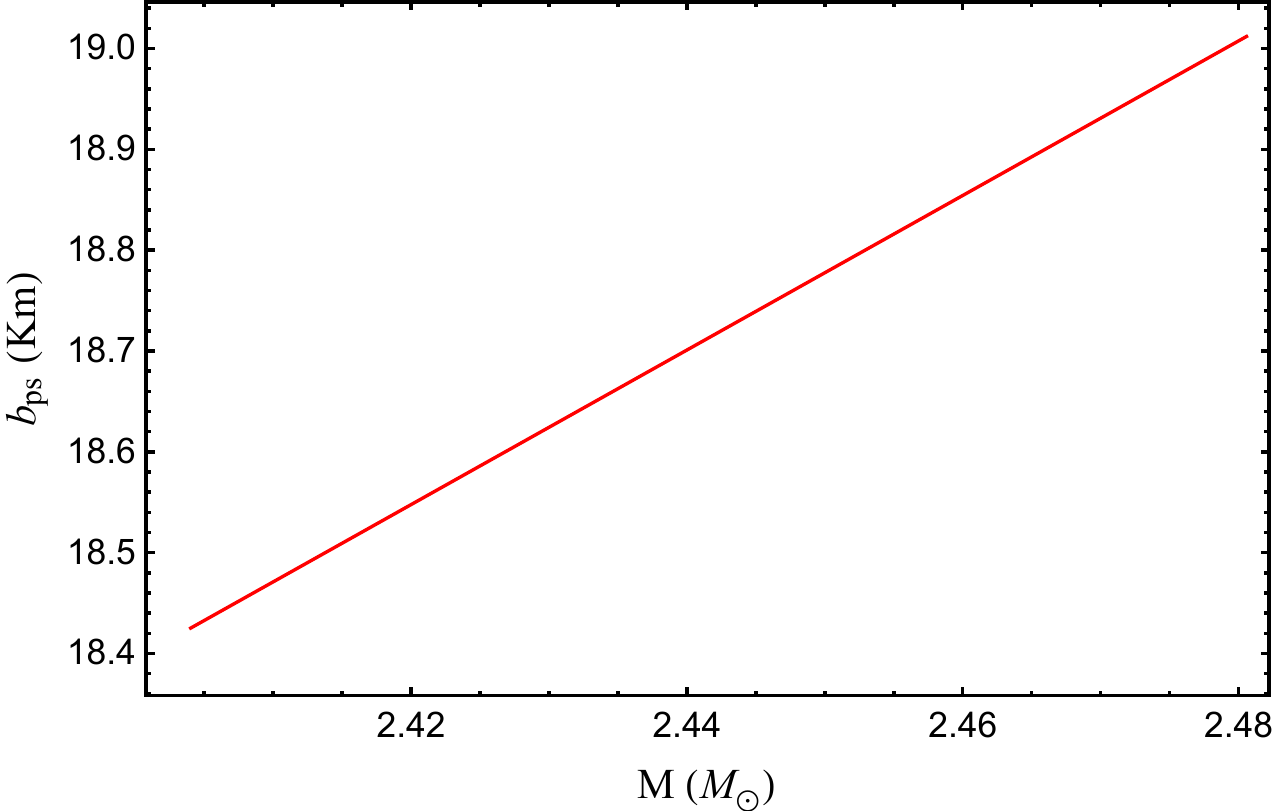}
		\subcaption{}
		\label{fig16}
	\end{minipage}
	\hfil
	\begin{minipage}{0.4\textwidth}
		\centering
		\includegraphics[width=8cm]{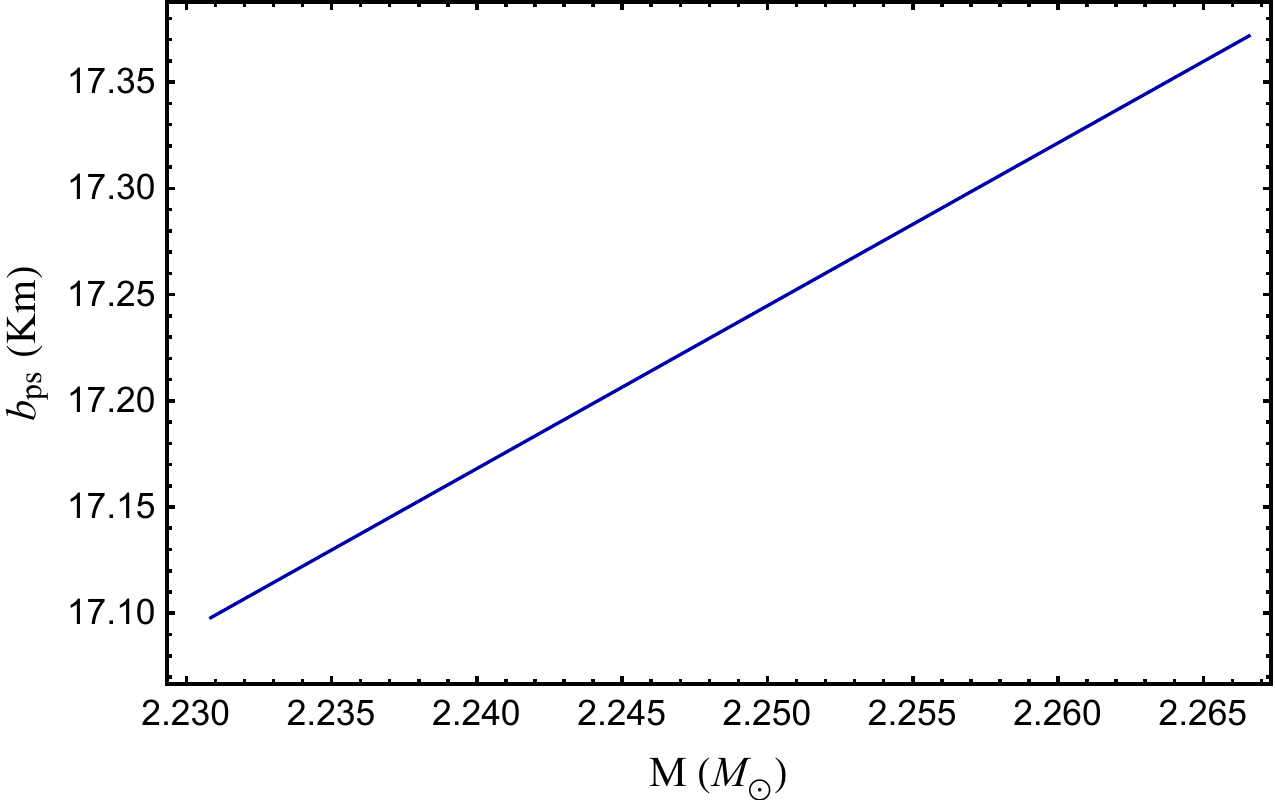}
		\subcaption{}
		\label{fig17}
	\end{minipage}
	\hfil
	\begin{minipage}{0.4\textwidth}
		\centering
		\includegraphics[width=8cm]{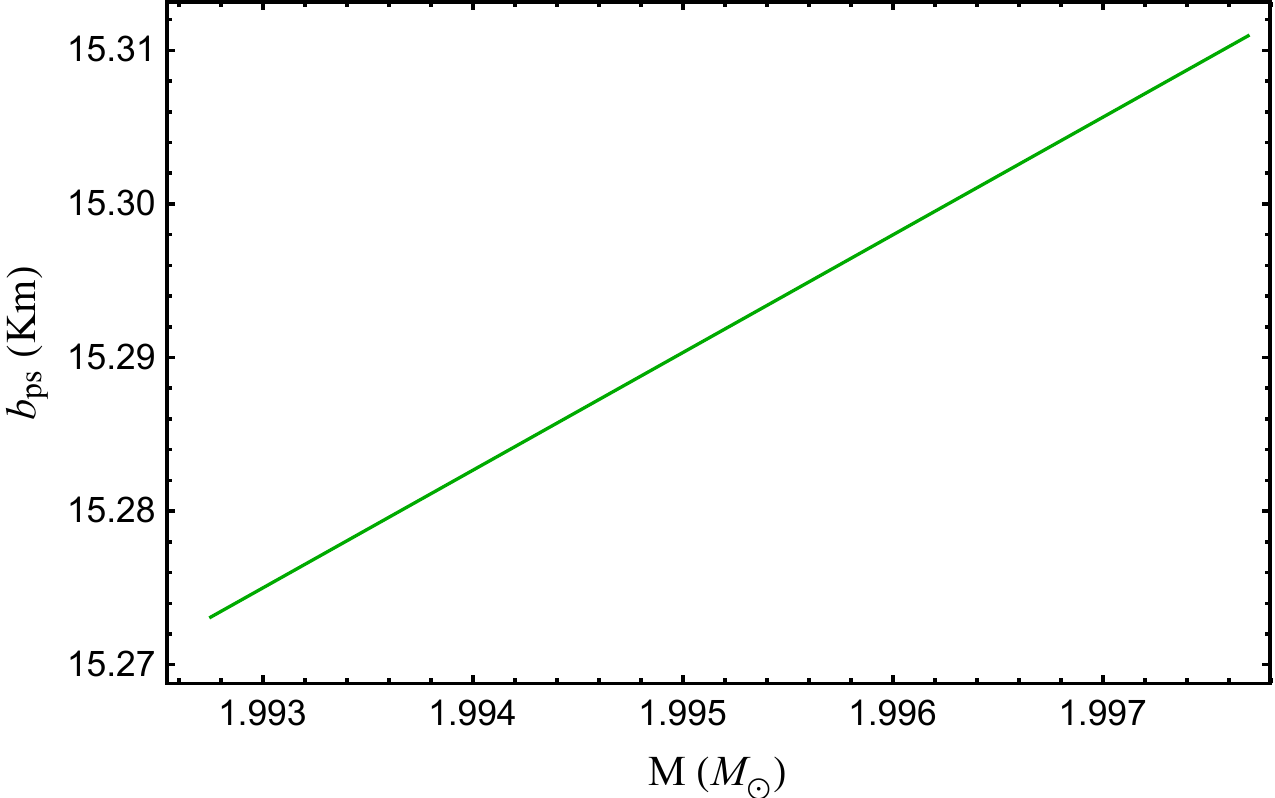}
		\subcaption{}
		\label{fig18}
	\end{minipage}
	\caption{\raggedright Shadow impact parameter vs mass plot for (a) $\alpha=5.0$ and $\beta=-0.98$, (b) $\alpha=5.0$ and $\beta=-0.83$, and (c) $\alpha=5.0$ and $\beta=-0.65$.}
\end{figure*}

For our echo-producing configurations where $R \le R_{\text{ps}} = 3M$, light rays grazing the stellar surface are captured or deflected by the photon sphere barrier, casting a distinct shadow-like optical feature (a "quark star shadow") surrounded by possible thin accretion emission. Using Eq.~(\ref{eq9}) and our calculated echo-specific mass-radius bounds, the effective shadow radius $b_{\text{ps}}$ provides a scale that can be differentiated from standard neutron star surface emission models. Future high-frequency EHT observations targeting galactic compact objects or post-merger remnants could constrain or falsify these ultra-compact configurations independently of GW detections. In the present parameter space, we have studied the variation of $b_{\text{ps}}$ with maximum mass, and the results are illustrated in Figs.~\ref{fig13}-\ref{fig18}.  


We observe that, for a fixed value of $\beta$ ($\alpha$), the maximum mass decreases with increasing $\alpha$ ($\beta$), accompanied by a corresponding contraction of the photon sphere impact parameter $b_{\text{ps}}$. This behaviour indicates that increasing either the quadratic-curvature contribution or the gravity-matter coupling modifies the exterior space-time geometry, leading to a suppression of the effective space-time curvature. As a consequence, the photon sphere moves inward, resulting in a smaller apparent shadow size. Further, we have obtained the values of $b_{\text{ps}}$ at the photon sphere crossing, and the corresponding results are tabulated in Tables~\ref{tab8} and \ref{tab9}.

\begin{table*}[t!]
		\centering
		\caption{\raggedright Tabulation of shadow impact parameter within the allowed parameter space. Here, panel (a) corresponds to $\beta=-0.5$ and $\alpha=10,~8.1 ~\text{and},~6.4$, respectively and panel (b) corresponds to $\alpha=5.0$ and $\beta=-0.98,~-0.83 ~\text{and} -0.65$, respectively.}
		\begin{minipage}{0.5\textwidth}
			\subcaption{}
			\label{tab8}
				\begin{tabular}{ccc}
				\hline
				\hline
				$\beta$ & $\alpha$ & $b_{\text{ps}}$ (Km) \\
				\hline
				\multirow{3}{*}{-0.5} & 10 & 19.02 \\
				& 8.1 & 17.13 \\
				& 6.4 & 15.22 \\ 
				\hline
				\hline
			\end{tabular}
		\end{minipage}
	\hfil
	\begin{minipage}{0.491\textwidth}
		\hspace{-2cm}
		\subcaption{}
		\label{tab9}
		\begin{tabular}{ccc}
			\hline
			\hline
			$\alpha$ & $\beta$ & $b_{\text{ps}}$ (Km) \\
			\hline
		    \multirow{3}{*}{5} & -0.98 & 19.01 \\
			& -0.83 & 17.37 \\
			& -0.65 & 15.31\\ 
			\hline
			\hline
		\end{tabular}
	\end{minipage}
	
\end{table*}
\subsection{Observational detectability in the next generation GW networks} Our obtained results indicate the strange stars satisfying hydrostatic equilibrium and photon sphere constraints produce characteristic echo frequencies in the kHz range. At present, the LVK detectors have optimal sensitivity in the 10-1000 Hz range. However, our predictions align directly with the planned decade upgrades for the KAGRA network \cite{Akutsu} operating in the kHz regime, as well as the next-generation ground-based interferometers such as the Einstein Telescope (ET) and Cosmic Explorer (CE). These observatories are specially designed to capture the post-merger sub-kHz and kHz bands and make the detection of high frequency trapped modes $(f_{\text{echo}})$ an attainable probe to distinguish $f(R+\alpha\,R^{2},T)$ gravity-strange stars from GR black holes. 
\subsection{Estimating the quasi-normal modes in the eikonal limit}
The study of quasi-normal modes (QNMs) was initially developed in the context of linear perturbations of the exterior Schwarzschild space-time \cite{Regge}. When a compact object is subjected to an external perturbation, QNMs are excited and manifest as damped oscillations in its response. These characteristic oscillations are determined by the underlying space-time geometry and, therefore, provide a valuable theoretical framework for probing the properties of compact objects within both GR and modified theories of gravity. In light of the GW observations by the LVK collaborations, the post-merger ringdown signal is generally expected to be dominated by QNMs \cite{Berti,Kokkotas1}. Consequently, the study of QNMs in post-merger systems and during the late stages of gravitational collapse can provide important insights into the nature and structure of compact objects and may help distinguish between different theoretical models of gravity and compact star configurations.

In the present study, we have noted that the obtained echo frequencies are in the kHz order, and hence, it is appropriate to impose the eikonal limit on the characterisation of QNMs. Following the work of Ferrari and Mashhoon \cite{Ferrari1}, we have noted that the spectrum of QNM is expressed as:
\begin{equation}
	\omega=\Bigg(\ell+\frac{1}{2}\Bigg)\Omega_{\text{ps}}-\iota\,\Bigg(n+\frac{1}{2}\Bigg)\lambda_{\text{ps}}, \label{eq15} 
\end{equation}  
where $\Omega_{\text{ps}}=\sqrt{\frac{M}{R_{\text{ps}}^{3}}}$, and $\lambda_{\text{ps}}=(3\sqrt{3}M)^{-1}$ is the Lyapunov exponent. Notably, the impact of the present parameter space is incorporated into the QNMs through the calculation of masses which vary with suitable choices of $\alpha$ and $\beta$. We have tabulated the variation of QNMs for the parametric choices of model parameters in Tables~\ref{tab10} and \ref{tab11}.

\begin{table*}
	\centering
	\caption{\raggedright QNM frequencies for different combinations of $\ell$ and $n$ within the allowed parameter space. Here, panel (a) corresponds to $\beta=-0.5$ and $\alpha=10,~8.1 ~\text{and},~6.4$, respectively and panel (b) corresponds to $\alpha=5.0$ and $\beta=-0.98,~-0.83 ~\text{and} -0.65$, respectively.}
	\begin{minipage}{0.45\textwidth}
		\subcaption{}
		\label{tab10}
		\begin{tabular}{ccccc}
			\hline
			\hline
			$\beta$  & $\alpha$  & $f (l=2,\,n=0)$ &
			$f(l=2,\,n=1)$ & $f(l=3,\,n=0)$ \\
			&&(kHz) & (kHz) & (kHz)  \\
			\hline
			\multirow{3}{*}{-0.5} & 10  & 6.24 & 6.24 & 8.74 \\
			& 8.1  & 6.93 & 6.93 & 9.70 \\ 
			& 6.4 & 7.80 & 7.80 & 10.92 \\
			\hline
			\hline
		\end{tabular}
	\end{minipage}
\hfil
		\begin{minipage}{0.45\textwidth}
		\subcaption{}
		\label{tab11}
		\begin{tabular}{ccccc}
			\hline
			\hline
			$\alpha$  & $\beta$  & $f (l=2,\,n=0)$ & $f(l=2,\,n=1)$ & $f(l=3,\,n=0)$ \\
			&&(kHz) & (kHz) & (kHz) \\
			\hline
			\multirow{3}{*}{5} & -0.98 & 6.24 & 6.24 & 8.74 \\
			& -0.83 & 6.83 & 6.83 & 9.57 \\ 
			& -0.65 & 7.75 & 7.75 & 10.86 \\
			\hline
			\hline
		\end{tabular}
		\end{minipage}
\end{table*}

We note that the eikonal frequencies increase with systematic increase of quadratic curvature parameter $(\alpha)$ and gravity-matter coupling $(\beta)$. This trends has several observational implications, namely (i) since the characteristic frequencies are shifted to the kHz regime, these modes are the primary targets for next-generation observatories (such as the ET and CE) and post-O5 KAGRA upgrades designed specifically for high-frequency sensitivity \cite{Akutsu}, (ii) the inverse relation between the eikonal frequency and the shadow impact parameter ($b_{\text{ps}} \sim 1/\Omega_{\text{ps}}$) implies that a stronger gravity-matter coupling shrinks the effective photon sphere. This provides an observational bridge between post-merger GW ringdowns and ultra-high-resolution optical shadow measurements from the EHT, and (iii) since modifications in $\alpha$ and $\beta$ increase the eikonal frequencies while maintaining the hydrostatic equilibrium, detecting a higher than usual QNM frequencies in the post-merger ringdown may act as a diagnostic tool to disentangle modified gravity effects from nuclear EoS uncertainties. 
\section{Conclusions}\label{sec4}
We have investigated the possibility of GW echoes from SS in the framework of $f(\tilde{R},T)=R+\alpha\,R^{2}+2\beta\,T$ gravity. To represent the interior matter sector of SS, we have considered the simple yet phenomenological MIT bag model EoS \cite{Kettner}. To reiterate, the emergence of echoes is driven by the existence of photon sphere. In GR, the photon sphere is established at R=3M. A compact star possesses a photon sphere and forms a GW trapping cavity, if its radius is less than 3M. In the present framework, we have noted that this photon sphere radius is also maintained in the vacuum limits. Pani and Ferrari \cite{Pani} showed that the occurrence of GW echoes requires the compactness of NS to be sufficiently close to the Buchdahl limit \cite{Buchdahl}, where the corresponding radius is given by $R_{b}=\frac{9M}{4}$. Hence, in GR, the compactness must lie within the range $1/3\leq\frac{M}{R}\leq4/9$ for the production of low-frequency GW echoes. However, in presence of modified gravity, the Buchdahl limit is extended in the following form, $\frac{M}{R}=\frac{4}{9}-\frac{a}{6}$ \cite{Burikham}, where $a=4\pi\,p_{eff}(R)R^{2}$. Hence, in the present context, the compactness must lie in the range, $\frac{1}{3}\leq\frac{M}{R}\leq\left(\frac{4}{9}-\frac{a}{6}\right)$ to generate GW echoes. 

Keeping this pre-requisites in mind, we have constructed and solved the modified TOV equations by employing the MIT bag model EoS within the framework of $f(\tilde{R},T)=R+\alpha\,R^{2}+2\beta\,T$ gravity. The bag parameter, $(B_{g})$, is varied within the stable strange matter range, i.e., $57.55\leq\,B_{g}\leq\,95.11~\mathrm{MeV/fm^{3}}$ \cite{Madsen}. To obtain ghost-free solutions, we have considered $f_{RR}\geq0$ which leads to $\alpha>0$ and the non-minimal matter coupling $(\beta)$ is varied up to the point that yields physically acceptable TOV solutions. Since the lower compactness threshold for the existence of a photon sphere is $\frac{1}{3}$, we present the $M$-$R$ relations only for configurations with compactness exceeding this limit. The corresponding results are shown in Figs.~\ref{fig1} and \ref{fig2} and summarised in Tables~\ref{tab1} and \ref{tab2}. 

In this regard, we have noted that $\beta$ must be negative to ensure $\frac{M}{R}>\frac{1}{3}$. An important outcome of the present analysis is revealed by the intersections of the photon-sphere condition with the $M$-$R$ curves. These intersection points identify the critical stellar configurations beyond which GW echoes become feasible. Configurations lying below these thresholds do not admit a photon sphere, indicating that the necessary condition for the production of GW echoes is not satisfied. Imposing the modified Buchdahl constraint~\cite{Burikham} in conjunction with the compactness requirement, $(\geq \frac{1}{3})$, excludes a subset of the TOV solutions from the physically viable parameter space. Consequently, the allowed stellar configurations are refined, resulting in modified values of the mass, radius, and compactness that are consistent with both the stability criterion and the conditions necessary for the existence of a photon sphere. The new parameter ranges are tabulated in Tables~\ref{tab3} and \ref{tab4}. Hence, from the considerations of hydrostatic equilibrium, the $M$-$R$ values of a SS follows Tables~\ref{tab1} and \ref{tab2}, however, to produce echoes, the $M$-$R$ values must satisfy the limits summarised in Tables~\ref{tab3} and \ref{tab4}. From the echo perspective, these $M$-$R$ values must be regarded as the new $M$-$R$ limits of SS. 

Under suitable variations of $\alpha$ and $\beta$, we have noted that with increasing $B_{g}$, the maximum mass decreases owing to the softer nature of the EoS. As $B_{g}$ increases, the difference between perturbative and non-perturbative vacua increases and consequently, the compressibility of the stellar configuration increases. These feature reduce the maximum mass of the SS. Within the chosen parameter space, we have shown the variations of compactness with maximum mass in Figs.~\ref{fig3} and \ref{fig4}. Figs.~\ref{fig3} and \ref{fig4} identify the compactness window allowed by the modified Buchdahl bound, within which the configurations can support GW echoes.

Having established that the newly obtained mass limits satisfy the compactness criterion for the existence of a photon sphere, we examine the corresponding GW echo properties. Since different combinations of $\alpha$, $\beta$, and $B_{g}$ yield distinct $M$-$R$ relations and compactness profiles, the associated GW echo characteristics also vary. Using Eq.~(\ref{eq12}), we compute the echo time for the maximum mass configurations, and the results are shown in Figs.~\ref{fig7} and \ref{fig8}. We find that the echo time decreases with increasing $B_{g}$, although the extent of this decrease depends on the chosen values of $\alpha$, $\beta$, and $B_{g}$ (see Tables~\ref{tab1} and \ref{tab2}). This behaviour can be attributed to the softening of the MIT bag model EoS with increasing $B_{g}$, which produces more compact stellar configurations and reduces the effective cavity traversed by the trapped GW signal. Consequently, the echo frequency increases with decreasing echo time, as illustrated in Figs.~\ref{fig9} and \ref{fig10}. The corresponding echo times and frequencies for the newly obtained maximum mass of SS configurations are summarised in Tables~\ref{tab5} and \ref{tab6}. Furthermore, we have studied the stability of the present model, using HZN condition, in Figs.~\ref{fig11} and \ref{fig12}. The incorporation of photon sphere constraint in the HZN criterion shows that the new maximum mass limits, obtained in this model, maintain the HZN condition, indicating a stable stellar structure.   

An important observational aspect of the present model is the simultaneous connection between the ultracompact structure, photon-sphere properties, and post-merger GW signatures. The allowed configurations exhibit shadow impact parameters of order $15$-$19$ Km, as described in Tables~\ref{tab8} and \ref{tab9}, and characteristic QNM frequencies in the kHz regime, as tabulated in Tables~\ref{tab10} and \ref{tab11}, with both quantities varying systematically with the parametric choices of model parameters, $\alpha$ and $\beta$. These results suggest that electromagnetic observations of compact object shadows and future high-frequency GW observations could provide complementary constraints on $f(R+\alpha R^{2},T)$ strange star configurations. While a direct comparison with EHT or future GW data requires a detailed treatment of source properties and detector sensitivity, the combined shadow and QNM signatures provide an observationally motivated avenue for testing the viability of these ultracompact configurations. We have also noted that the echo frequencies are in the kHz order, which is particularly interesting in the context of ten-year upgrade roadmap for KAGRA for the era of post-O5 GW astronomy \cite{Akutsu}. KAGRA collaboration is planning for an upgrade to $\sim$kHz range, which is sensitive to binary NS coalescence. Furthermore, detection of the signals in the kHz range would significantly enhance the precision in the measurement of tidal properties in the post-merger phase. Hence, the present framework provides a unified approach to investigate post-merger GW echoes from compact binary coalescences. The detection of echoes within the obtained frequency range may provide observational evidence for SS in quadratic curvature-matter coupled gravity, while the existence of a photon sphere and the associated GW echoes place tight constraints on their maximum mass. We have summarised the novelty of the obtained results in comparison to the previous studies on GW echoes from SS/compact stars in Table~\ref{tab7}. The existence of a photon sphere and the modified Buchdahl bound define a new $M$-$R$ limit for SS. 
\begin{table*}[t!]
	\centering
	\caption{Comparison of the present work with previous studies on GW echoes from compact stars.}
	\label{tab7}
	\begin{tabular}{lp{5.6cm}p{7.2cm}}
		\hline
		\hline
		Previous work & Main focus & Novel aspect of the present work \\
		\hline
		Urbano \& Veerm\"ae \cite{Urbano} &
		Investigated GW echoes from ultracompact exotic stars and the associated echo phenomenology. &
		Rather than assuming generic exotic compact objects, this work constructs physically self-consistent SS configurations from modified TOV equations and directly relates the echo properties to the stellar EoS and gravity parameters. \\ \hline
		Mannarelli \& Tonelli \cite{Mannarelli} &
		Studied GW echoes from SS in GR using the MIT bag model EoS. &
		Shows that modified gravity substantially alters the stellar structure, allowing more compact SS, revising the maximum mass, and changing the echo properties beyond the GR predictions. \\ \hline
		Pani \& Ferrari \cite{Pani} &
		Established the conditions for GW echoes from ultracompact NS in GR, highlighting the importance of the photon sphere and the Buchdahl limit. &
		Extends the analysis to quadratic curvature-matter coupled gravity, derives the modified TOV equations, employs the modified Buchdahl bound, and demonstrates that modified gravity enables stable SS to satisfy the compactness condition required for GW echoes. \\ \hline
		C. Zhang \cite{Zhang1} &
		Studied GW echoes from interacting quark stars in GR. &
		Investigates GW echoes in a modified gravity framework, providing the combined effects of the bag constant, quadratic curvature correction ($\alpha$), and matter-geometry coupling ($\beta$) on the stellar structure and GW echoes. \\ \hline		
		Bora \& Goswami \cite{Bora}&
		Analysed radial oscillations and GW echoes for SS using different EoS in GR. &
		Instead of comparing EoSs alone, the present work demonstrates how quadratic curvature corrections and matter-geometry coupling modify the compactness threshold, photon-sphere formation, and echo frequencies. \\ \hline
		Bora et al. \cite{Bora1}&
		Investigated SS and GW echoes in Palatini $f(R)$ gravity, showing that modified gravity can increase stellar compactness and produce echoes. &
		Introduces a fundamentally different gravity model containing both quadratic curvature corrections and explicit matter-geometry coupling. The additional coupling parameter $\beta$ provides an extra degree of freedom that modifies the stellar equilibrium, compactness and echo characteristics.\\ \hline
		Bora \& Goswami \cite{Bora2}&
		Studied GW echoes from compact stars in metric $f(R,T)$ gravity using MIT bag and CFL EoSs, and constrained different parameters from the existence of echoes. & The present work extends beyond linear $f(R,T)$ gravity by incorporating the quadratic curvature term $\alpha R^2$, allowing the combined influence of higher-order curvature and matter coupling on the stellar structure, photon sphere and GW echoes to be quantified. Moreover, the analysis derives revised maximum mass-radius limits associated specifically with echo-producing SS.\\ \hline
		Present work &
		GW echoes in quadratic curvature-matter coupled gravity with MIT bag model SS. &
		(i) Derives and solves the modified TOV equations for $f(R+\alpha R^{2},T)$ gravity, (ii) establishes the parameter space that admits photon spheres, (iii) identifies the intersection of the photon-sphere condition with the $M$-$R$ curve as the onset of GW echo producing configurations, (iv) derives new $M$-$R$ limits specifically for echo-producing SS by simultaneously imposing hydrostatic equilibrium and the modified Buchdahl constraint, (v) predicts kHz GW echo frequencies and demonstrates their dependence on $\alpha$, $\beta$, and the bag constant,
		(vi) the obtained frequency band is particularly interesting from the perspective of KAGRA upgradation \cite{Akutsu}, (vii) proposes GW echoes as a probe of both modified gravity and the maximum mass of SS, (viii) the stability of mass points are analysed from HZN condition, (ix) to assess the observational signatures from the model, optical shadows, EHT/LVK constraints, and QNM frequencies are analysed. \\ \hline
		\hline
	\end{tabular}
\end{table*}

The present analysis can be extended in several directions. A natural next step is to investigate rotating SS within the framework of quadratic curvature-matter coupled gravity, since astrophysical compact stars are expected to possess non-negligible angular momentum. Rotation modifies the stellar compactness, photon-sphere properties, and GW echo characteristics, thereby providing a more realistic description of post-merger remnants. It would also be worthwhile to explore more sophisticated EoSs incorporating colour superconductivity, finite-temperature effects, and strong magnetic fields to assess the robustness of the present results. On the observational side, the predicted echo frequencies may be confronted with future high-frequency GW observations from next-generation detectors, particularly upgraded KAGRA and other planned observatories operating in the kHz band. Such studies could provide stringent constraints on both the existence of SS and the viability of quadratic curvature-matter coupled gravity, offering an independent probe of strong-field gravity beyond GR. In this study, we identify the region of the $(\alpha,\beta,B_{g})$ parameter space in which stable strange-star configurations can simultaneously satisfy the photon-sphere and modified Buchdahl constraints. The resulting allowed mass-radius values are explicitly tabulated in Tables~\ref{tab3} and \ref{tab4} of the manuscript. A further outcome that distinguishes the present analysis is the direct connection between the modified stellar structure and the echo observables. We find that variations of $\alpha,\beta$, and the bag constant, $B_{g}$, modify the stellar compactness, echo time, and echo frequency. Within the allowed parameter space, the predicted echo frequencies lie in the kHz range, with the maximum-mass configurations satisfying the echo constraints. Thus, the analysis establishes a direct chain $(\alpha, \beta, B_{g})$ $\rightarrow$ (mass, radius, compactness) $\rightarrow$ photon sphere $\rightarrow$ $\tau_{\text{echo}}$, $f_{\text{echo}}$, rather than treating the echo properties independently of the stellar equilibrium solution. The corresponding numerical results are presented in Tables~\ref{tab5} and \ref{tab6}. Table~\ref{tab7} has been expanded as a concise comparison showing that the present work differs from earlier studies by combining quadratic curvature corrections and explicit matter-gravity coupling, deriving the corresponding self-consistent stellar structure, imposing the modified Buchdahl and photon-sphere constraints, and obtaining echo-specific mass-radius limits and kHz echo frequencies. The central result is the derivation of new, echo-specific constraints on strange star configurations arising from the interplay between the modified gravitational dynamics, the strange matter EoS, the photon-sphere requirement, and the modified Buchdahl bound. These constraints provide an independent route to limiting the maximum mass of strange stars, beyond the usual requirement of hydrostatic equilibrium alone.
\section*{Acknowledgements} DB is thankful to the Department of Science and Technology (DST), Govt. of India, for providing the fellowship vide no:  DST/INSPIRE Fellowship/2021/IF210761. PKC gratefully acknowledges support from IUCAA, Pune, India under Visiting Associateship programme. The work of Kazuharu Bamba was supported in part by the JSPS KAKENHI Grants No. 24KF0100 and No. 25KF0176, and a grant-in-aid of academic research of the Yamaguchi Scholarship Foundation.


\begin{thebibliography}{99}
	\bibitem{Abbott} B. P. Abbott et al. (LIGO Scientific, Virgo), Observation of Gravitational Waves from a Binary Black Hole Merger, Phys. Rev. Lett. {\bf 116}, 061102 (2016) \href{arXiv:1602.03837 [gr-qc]}{arXiv:1602.03837 [gr-qc]}.
	\bibitem{Abbott1} B. P. Abbott et al. (LIGO Scientific, Virgo), GWTC-1: A Gravitational-Wave Transient Catalog of Compact Binary
	Mergers Observed by LIGO and Virgo during the First and Second Observing Runs, Phys. Rev. X {\bf 9}, 031040 (2019),
	\href{arXiv:1811.12907 [astro-ph.HE]}{arXiv:1811.12907 [astro-ph.HE]}.
	\bibitem{Abbott2} R. Abbott et al. (LIGO Scientific, Virgo), GWTC-2: Compact Binary Coalescences Observed by LIGO and Virgo During the First Half of the Third Observing Run, Phys. Rev. X {\bf 11}, 021053 (2021), \href{arXiv:2010.14527 [gr-qc]}{arXiv:2010.14527 [gr-qc]}.
	\bibitem{Abbott3} R. Abbott et al. (KAGRA, VIRGO, LIGO Scientific), GWTC-3: Compact Binary Coalescences Observed by LIGO and
	Virgo during the Second Part of the Third Observing Run, Phys. Rev. X {\bf 13}, 041039 (2023), \href{arXiv:2111.03606 [gr-qc]}{arXiv:2111.03606 [gr-qc]}.
	\bibitem{Abac} A. G. Abac et al. (LIGO Scientific, VIRGO, KAGRA), GWTC-4.0: Updating the Gravitational-Wave Transient Catalog
	with Observations from the First Part of the Fourth LIGO-Virgo-KAGRA Observing Run, Astrophys. J. Lett. {\bf 1004}, L22 (2026), \href{arXiv:2508.18082 [gr-qc]}{arXiv:2508.18082 [gr-qc]}.
	\bibitem{Abac1} A. G. Abac et al. (LIGO Scientific, Virgo, KAGRA), GW250114: Testing Hawking’s Area Law and the Kerr Nature of
	Black Holes, Phys. Rev. Lett. {\bf 135}, 111403 (2025), \href{arXiv:2509.08054 [gr-qc]}{arXiv:2509.08054 [gr-qc]}.
	\bibitem{Abac2} A. G. Abac et al. (LIGO Scientific, Virgo, KAGRA), Black Hole Spectroscopy and Tests of General Relativity with
	GW250114,Phys. Rev. Lett. {\bf 136}, 041403 (2026), \href{arXiv:2509.08099 [gr-qc]}{arXiv:2509.08099 [gr-qc]}.
	\bibitem{Isi} M. Isi et al., Testing the no-hair theorem
	with GW150914, Phys. Rev. Lett. {\bf 123}, 111102 (2019), \href{arXiv:1905.00869 [gr-qc]}{arXiv:1905.00869 [gr-qc]}.
	\bibitem{Capano} C. D. Capano et al., Multimode Quasinormal Spectrum from a Perturbed Black Hole, Phys. Rev. Lett. {\bf 131}, 221402 (2023), \href{arXiv:2105.05238 [gr-qc]}{arXiv:2105.05238 [gr-qc]}.
	\bibitem{Cotesta} R. Cotesta, G. Carullo, E. Berti, and V. Cardoso, Analysis of Ringdown Overtones in GW150914,	Phys. Rev. Lett. {\bf 129}, 111102 (2022), \href{arXiv:2201.00822 [gr-qc]}{arXiv:2201.00822 [gr-qc]}.
	\bibitem{Cardoso} V. Cardoso, E. Franzin, and P. Pani, Is the gravitational-wave ringdown a probe of the event horizon?
	Phys. Rev. Lett. {\bf 116}, 171101 (2016), [Erratum: Phys.Rev.Lett. {\bf 117}, 089902 (2016)], \href{arXiv:1602.07309 [gr-qc]}{arXiv:1602.07309 [gr-qc]}.
	\bibitem{Cardoso1} V. Cardoso et al., Gravitational-wave signatures of exotic compact objects and of quantum corrections at the horizon scale, Phys. Rev. D {\bf 94}, 084031 (2016), \href{arXiv:1608.08637 [gr-qc]}{arXiv:1608.08637 [gr-qc]}.
	\bibitem{Bueno} P. Bueno et al., Echoes of Kerr-like wormholes,
	Phys. Rev. D {\bf 97}, 024040 (2018), \href{arXiv:1711.00391 [gr-qc]}{arXiv:1711.00391 [gr-qc]}.
	\bibitem{Zhang} J. Zhang and S.-Y. Zhou, Can the graviton have a large mass near black holes? Phys. Rev. D {\bf 97}, 081501
	(2018), \href{arXiv:1709.07503 [gr-qc]}{arXiv:1709.07503 [gr-qc]}.
	\bibitem{Urbano} A. Urbano and H. Veerm\"ae, On gravitational echoes from ultracompact exotic stars, J. Cosmol. Astropart. Phys. {\bf 04}, 011 (2019), \href{arXiv:1810.07137 [gr-qc]}{arXiv:1810.07137 [gr-qc]}.
	\bibitem{Mannarelli} M. Mannarelli and F. Tonelli, Gravitational wave echoes from strange stars, Phys. Rev. D {\bf 97}, 123010 (2018), \href{arXiv:1805.02278 [gr-qc]}{arXiv:1805.02278 [gr-qc]}.
	\bibitem{Li} Z.-P. Li and Y.-S. Piao, Mixing of gravitational wave echoes, Phys. Rev. D {\bf 100}, 044023 (2019), \href{arXiv:1904.05652[gr-qc]}{arXiv:1904.05652[gr-qc]}.
	\bibitem{Dong} R. Dong and D. Stojkovic, Gravitational wave echoes from black holes in massive gravity, Phys. Rev. D {\bf 103}, no.2, 024058 (2021) \href{arXiv:2011.04032 [gr-qc]}{arXiv:2011.04032 [gr-qc]}.
	\bibitem{Weinberg} S. Weinberg, {\it Gravitation and Cosmology} (John Wiley and
	Sons, NewYork, 1972).
	\bibitem{Misner} C. W. Misner, K. S. Thorne, and J. A. Wheeler, {\it Gravitation} (W.H. Freeman, San Francisco, 1973).
	\bibitem{Claudel} C. -M. Claudel, K. S. Virbhadra, and G. F. R. Ellis, The geometry of photon surfaces, J. Math. Phys. {\bf 42}, 818 (2001), \href{arXiv:gr-qc/0005050[gr-qc]}{arXiv:gr-qc/0005050[gr-qc]}.
	\bibitem{Shapiro} S. L. Shapiro and S. A. Teukolsky, {\it Black holes,white dwarfs, and neutron stars: The physics of compact objects} (1983). 
	\bibitem{Iyer} B. R. Iyer, C. V. Vishveshwara, and S. V. Dhurandhar, Ultracompact $(R<3 M)$ objects in general relativity, Class. Quantum Gravity {\bf 2}, 219 (1985).
	\bibitem{Nemiroff} R. J. Nemiroff, P. A. Becker, and K. S. Wood, Properties of ultracompact neutron stars, Astrophys. J. {\bf 406}, 590 (1993).
	\bibitem{Ferrari} V. Ferrari and K. D. Kokkotas, Scattering of particles by neutron stars: Time-evolutions for axial perturbations, Phys. Rev. D {\bf 62}, 107504 (2000), \href{arXiv:gr-qc/0008057[gr-qc]}{arXiv:gr-qc/0008057[gr-qc]}.
	\bibitem{Abbott4} B. Abbott et al. (LIGO Scientific Collaboration and Virgo Collaboration), GW170817: Observation of Gravitational Waves from a Binary Neutron Star Inspiral, Phys. Rev. Lett. {\bf 119}, 161101 (2017), \href{arXiv:1710.05832[gr-qc]}{arXiv:1710.05832[gr-qc]}.
	\bibitem{Abedi} J. Abedi and N. Afshordi, Echoes from the Abyss: A highly spinning black hole remnant for the binary neutron star merger GW170817, J. Cosmol. Astropart. Phys. {\bf 1911}, 010 (2019) \href{arXiv:1803.10454[gr-qc]}{arXiv:1803.10454[gr-qc]}.
	\bibitem{Pani} P. Pani and V. Ferrari, On gravitational-wave echoes from neutron-star binary coalescences, Class. Quantum Gravity {\bf 35}, 15LT01 (2018) \href{arXiv:1804.01444}{arXiv:1804.01444}. 
	\bibitem{Buchdahl} H. A. Buchdahl, General Relativistic Fluid Spheres, Phys. Rev. {\bf 116}, 1027 (1959).
	\bibitem{Itoh} N. Itoh, Hydrostatic Equilibrium 
	of Hypothetical Quark Stars, Prog. Theor. Phys. {\bf 44}, 291 (1970). 
	\bibitem{Madsen} J. Madsen, Physics and astrophysics of strange quark matter, Lect. Notes Phys. {\bf 516}, 162 (1999) \href{arXiv:astro-ph/9809032}{arXiv:astro-ph/9809032}.
	\bibitem{Bodmer} A. R. Bodmer, Collapsed Nuclei, Phys. Phys. Rev. {\bf 4}, 1601 (1971). 
	\bibitem{Witten} E. Witten, Cosmic separation of phases, Phys. Rev. D {\bf 30}, 272 (1984).
	\bibitem{Alford} M. Alford, Color superconducting quark matter, Ann. Rev. Nucl. Part. Sci. {\bf 51}, 131 (2001) \href{arXiv:hep-ph/0102047}{arXiv:hep-ph/0102047}.
	\bibitem{Alcock} C. Alcock, E. Farhi, and A. Olinto, STRANGE STARS, Astrophys. J. {\bf 310}, 261 (1986).
	\bibitem{Haensel} P. Haensel, J. Zdunik, and R. Schaeffer, Strange quark stars, Astron. Astrophys. {\bf 160}, 121 (1986).
	\bibitem{Kettner} Ch. Kettner et al., Structure and stability of strange and charm stars at finite temperatures, Phys. Rev. D {\bf 51}, 1440 (1995). 
	\bibitem{Zhang1} C. Zhang, Gravitational wave echoes from interacting quark stars, Phys. Rev. D {\bf 104}, 083032 (2021) \href{arXiv:2107.09654 [hep-ph]}{arXiv:2107.09654 [hep-ph]}.
	\bibitem{Bora} J. Bora and U.D. Goswami, Radial Oscillations and Gravitational Wave Echoes of Strange Stars for Various Equations of State, Mon. Not. R. Astron. Soc. {\bf 502}, 1557 (2021) \href{arXiv:2007.06553 [gr-qc]}{arXiv:2007.06553 [gr-qc]}.
	\bibitem{Bhattacharjee} D. Bhattacharjee, P. K. Chattopadhyay, K. Bamba, Maximum mass limit of strange stars in quadratic curvature-matter coupled gravity, Eur. Phys. J. C {\bf 86}, 389 (2026) \href{arXiv:2508.10524[gr-qc]}{arXiv:2508.10524[gr-qc]}.
	\bibitem{Tolman} R.C. Tolman, Static solutions of Einstein's field equations for spheres of fluid, Phys. Rev. {\bf 55}, 364 (1939).
	\bibitem{Oppenheimer} J.R. Oppenheimer and G.M. Volkoff, On massive neutron cores, Phys. Rev. {\bf 55}, 374 (1939).
	\bibitem{Starobinsky} A. A. Starobinsky, A new type of isotropic cosmological models without singularity, Phys. Lett. B {\bf 91}, 99 (1980) \href{https://doi.org/10.1016/0370-2693(80)90670-X}{https://doi.org/10.1016/0370-2693(80)90670-X}. 
	\bibitem{Akrami} Planck Collaboration: Y. Akrami et al., Planck 2018 results. X. Constraints on inflation. Astron. Astrophys. {\bf 641}, A10 (2020) \href{arXiv:1807.06211 [astro-ph.CO]}{arXiv:1807.06211 [astro-ph.CO]}.
	\bibitem{Feola} P. Feola, et al., The mass-radius relation for neutron stars in $f(R)=R+\alpha R^{2}$ gravity: a comparison between purely metric and torsion formulations. Phys. Rev. D {\bf 101}, 044037 (2020) \href{arXiv:1909.08847 [astro-ph.HE]}{arXiv:1909.08847 [astro-ph.HE]}.
	\bibitem{Capozziello2} S. Capozziello et al., The Mass-Radius relation for Neutron Stars in $f(R)$ gravity. Phys. Rev. D {\bf 93}, 023501 (2016) \href{arXiv:1509.04163 [gr-qc]}{arXiv:1509.04163 [gr-qc]}.
	\bibitem{Sotiriou} T. P. Sotiriou and V. Faraoni, $f(R)$ theories of gravity, Rev. Mod. Phys. {\bf 82}, 451 (2010) \href{arXiv:0805.1726 [gr-qc]}{arXiv:0805.1726 [gr-qc]}.
	\bibitem{Felice} A.~De Felice and S.~Tsujikawa, $f(R)$ theories, Living Rev. Rel. {\bf 13}, 3 (2010) \href{arXiv:1002.4928 [gr-qc]}{arXiv:1002.4928 [gr-qc]}.
	\bibitem{Nojiri} S.~Nojiri and S.~D.~Odintsov, Unified cosmic history in modified gravity: from $F(R)$ theory to Lorentz non-invariant models, Phys. Rept. {\bf 505}, 59 (2011) \href{arXiv:1011.0544 [gr-qc]}{arXiv:1011.0544 [gr-qc]}.
	\bibitem{Cai} Y.~F.~Cai, S.~Capozziello, M.~De Laurentis and E.~N.~Saridakis, $f(T)$ teleparallel gravity and cosmology, Rept. Prog. Phys. {\bf 79}, 106901 (2016) \href{arXiv:1511.07586 [gr-qc]}{arXiv:1511.07586 [gr-qc]}.
	\bibitem{Heisenberg} L.~Heisenberg, Review on $f(Q)$ gravity, Phys. Rept. {\bf 1066}, 1 (2024) \href{arXiv:2309.15958 [gr-qc]}{arXiv:2309.15958 [gr-qc]}.
	\bibitem{Nojiri1} S.~Nojiri, S.~D.~Odintsov and V.~K.~Oikonomou, Modified Gravity Theories on a Nutshell: Inflation, Bounce and Late-time Evolution, Phys. Rept. \textbf{692}, 1 (2017) \href{arXiv:1705.11098 [gr-qc]}{arXiv:1705.11098 [gr-qc]}. 
	\bibitem{Capozziello1} S.~Capozziello and M.~De Laurentis, Extended Theories of Gravity, Phys. Rept. {\bf 509}, 167-321 (2011) \href{arXiv:1108.6266 [gr-qc]}{arXiv:1108.6266 [gr-qc]}.
	\bibitem{Clifton} T.~Clifton, P.~G.~Ferreira, A.~Padilla and C.~Skordis, Modified Gravity and Cosmology. Phys. Rept. {\bf 513}, 1 (2012) \href{arXiv:1106.2476 [astro-ph.CO]}{arXiv:1106.2476 [astro-ph.CO]}.
	\bibitem{Bahamonde} S.~Bahamonde, K.~F.~Dialektopoulos, C.~Escamilla-Rivera, G.~Farrugia, V.~Gakis, M.~Hendry, M.~Hohmann, J.~Levi Said, J.~Mifsud and E.~Di Valentino, Teleparallel gravity: from theory to cosmology, Rept. Prog. Phys. {\bf 86}, no.2, 026901 (2023) \href{arXiv:2106.13793 [gr-qc]}{arXiv:2106.13793 [gr-qc]}.
	\bibitem{Sahni} V. Sahni and A. A. Starobinsky, The Case for a Positive Cosmological Lambda-term, Int. J. Mod. Phys. D {\bf 9}, 373 (2000) \href{arXiv:astro-ph/9904398}{arXiv:astro-ph/9904398}.
	\bibitem{Joyce} A. Joyce, L. Lombriser, and F. Schmidt, Dark Energy vs. Modified Gravity, Annu. Rev. Nucl. Part. Sci. {\bf 66}, 95 (2016) \href{arXiv:1601.06133 [astro-ph.CO]}{arXiv:1601.06133 [astro-ph.CO]}.
	\bibitem{Peebles} P. J. E. Peebles and B. Ratra, The Cosmological Constant and Dark Energy, Rev. Mod. Phys. {\bf 75}, 559 (2003) \href{arXiv:astro-ph/0207347}{arXiv:astro-ph/0207347}.
	\bibitem{Paddy} T. Padmanabhan, Cosmological Constant - the Weight of the Vacuum, Phys. Rep. {\bf 380}, 235 (2003) \href{arXiv:hep-th/0212290}{arXiv:hep-th/0212290}. 
	\bibitem{Astashenok1} A.V. Astashenok, S.D. Odintsov and V.K. Oikonomou, Chandrasekhar Mass Limit of White Dwarfs in Modified Gravity, Symmetry {\bf 15} (6), 1141 (2023) \href{arXiv:2211.14892 [gr-qc]}{arXiv:2211.14892 [gr-qc]}.
	\bibitem{Astashenok7} A.V. Astashenok, S. Capozziello and S.D. Odintsov, Extreme neutron stars from Extended Theories of Gravity, J. Cosmol. Astropart. Phys. {\bf 01}, 001 (2015) \href{arXiv:1408.3856 [gr-qc]}{arXiv:1408.3856 [gr-qc]}.
	\bibitem{Astashenok8} A.V. Astashenok, S. Capozziello and S.D. Odintsov, Maximal neutron star mass and the resolution of the hyperon puzzle in modified gravity, Phys. Rev. D {\bf 89} (10), 103509 (2014) \href{arXiv:1401.4546 [gr-qc]}{arXiv:1401.4546 [gr-qc]}.	
	\bibitem{Harko} T. Harko, F. S. N. Lobo, S. Nojiri, and S. D. Odintsov, $f(R,T)$ gravity. Phys. Rev. D {\bf 84}, 024020 (2011) \href{arXiv:1104.2669 [gr-qc]}{arXiv:1104.2669 [gr-qc]}.
	\bibitem{Bertolami} O. Bertolami, C. G. B\"ohmer, T. Harko, and F. S. N. Lobo, Extra force in $f(R)$ modified theories of gravity. Phys. Rev. D {\bf 75}, 104016 (2007) \href{arXiv:0704.1733 [gr-qc]}{arXiv:0704.1733 [gr-qc]}.
	\bibitem{Buchdahl1} H. A. Buchdahl, Non-Linear Lagrangians and Cosmological Theory, Mon. Not. R. Astron. Soc. {\bf 150}, 1 (1970) \href{https://doi.org/10.1093/mnras/150.1.1}{https://doi.org/10.1093/mnras/150.1.1}.
	\bibitem{Pretel} J. M. Z. Pretel, S. E. Jor\'as, R. R. R. Reis, and J. D. V. Arba\~nil, Radial oscillations and stability of compact stars in $f(R,T)=R+2\beta T$ gravity. J. Cosmol. Astropart. Phys. {\bf 04}, 064 (2021) \href{arXiv:2012.03342 [gr-qc]}{arXiv:2012.03342 [gr-qc]}.
	\bibitem{Burikham} P. Burikham, T. Harko and M. J. Lake, Mass bounds for compact spherically symmetric objects in generalized gravity theories, Phys. Rev. D {\bf 94}, 064070 (2016) \href{arXiv:1606.05515 [gr-qc]}{arXiv:1606.05515 [gr-qc]}.
	\bibitem{Cardoso2} V. Cardoso and P. Pani, Tests for the existence of black holes through gravitational wave echoes, Nat. Astron. {\bf 1}, 586 (2017) \href{arXiv:1709.01525 [gr-qc]}{arXiv:1709.01525 [gr-qc]}. 
	\bibitem{Mark} Z. Mark, A. Zimmerman, S. M. Du, and Y. Chen, A recipe for echoes from exotic compact objects, Phys. Rev. D {\bf 96}, 084002 (2017) \href{arXiv:1706.06155 [gr-qc]}{arXiv:1706.06155 [gr-qc]}.
	\bibitem{Kokkotas} K. D. Kokkotas, {\it Relativistic gravitation and gravitational radiation. Proceedings, School of Physics, Les Houches, France, September 26-October 6, 1995} (1995) pp. 89–102, \href{arXiv:gr-qc/9603024 [gr-qc]}{arXiv:gr-qc/9603024 [gr-qc]}.
	\bibitem{Andersson} N. Andersson, Y. Kojima and K. D. Kokkotas, On the oscillation spectra of ultracompact stars: An extensive survey of gravitational-wave modes, 
	Astrophys. J. {\bf 462}, 855 (1996) \href{arXiv:gr-qc/9512048 [gr-qc]}{arXiv:gr-qc/9512048 [gr-qc]}.
	\bibitem{Zeldovich} Ya B. Zel'dovich and I. D. Novikov, \textit{Relativistic Astrophysics, Stars and Relativity} \textbf{1}, (University of Chicago Press, Chicago, 1971).
	\bibitem{EHT} The Event Horizon Telescope Collaboration, 
	First M87 Event Horizon Telescope Results. I. The Shadow of the Supermassive Black Hole, Astrophys. J. {\bf 875}, L1 (2019), \href{arXiv:1906.11238 [astro-ph.GA]}{arXiv:1906.11238 [astro-ph.GA]}. 
	\bibitem{Akutsu} T. Akutsu et al. (KAGRA Collaboration), Decadal upgrade strategy for KAGRA toward post-O5 gravitational-wave astronomy, \href{arXiv:2508.03392 [gr-qc]}{arXiv:2508.03392 [gr-qc]}.
	\bibitem{Regge} T. Regge and J. A. wheeler, Stability of a Schwarzschild Singularity, Phys. Rev. {\bf 108}, 1063 (1957), \href{https://doi.org/10.1103/PhysRev.108.1063}{https://doi.org/10.1103/PhysRev.108.1063}.
	\bibitem{Berti} E. Berti, V. Cardoso and A. O. Starinets, Quasinormal modes of black holes and black branes, Class. Quantum Gravity {\bf 26}, 163001 (2009), \href{arXiv:0905.2975 [gr-qc]}{arXiv:0905.2975 [gr-qc]}.
	\bibitem{Kokkotas1} K. D. Kokkotas and B. G. Schmidt, Quasi-Normal Modes of Stars and Black Holes, Living Rev. Rel. {\bf 2}, 2 (1999), \href{arXiv:gr-qc/9909058}{arXiv:gr-qc/9909058}.
	\bibitem{Ferrari1} V. Ferrari and B. Mashhoon, New approach to the quasinormal modes of a black hole, Phys. Rev. D {\bf 30}, 295 (1984).
	\bibitem{Bora1} J. Bora, D. J. Gogoi and U. D. Goswami, Strange stars in $f(R)$ gravity palatini formalism and gravitational wave echoes from them, J. Cosmol. Astropart. Phys. {\bf 09}, 057 (2022) \href{arXiv:2204.05473 [gr-qc]}{arXiv:2204.05473 [gr-qc]}.
	\bibitem{Bora2} J. Bora and U. D. Goswami, Gravitational wave echoes from compact stars in $f(R,T)$ gravity, Phys. Dark Universe {\bf 38}, 101132 (2022) \href{arXiv:2207.12847 [gr-qc]}{arXiv:2207.12847 [gr-qc]}. 
\end{thebibliography}
\end{document}